\documentclass[sigconf]{acmart}

\usepackage{booktabs} 
\usepackage{multirow}
\usepackage{graphics}
\usepackage{graphicx}
\usepackage{algorithm}
\usepackage{algorithmic}
\usepackage{amsfonts}
\usepackage{amsmath}
\usepackage{color, url}
\usepackage[skip=0pt]{caption}
\usepackage[skip=0pt]{subcaption}
\usepackage{tabto}
\usepackage{url}
\usepackage{xcolor, colortbl}
\usepackage{booktabs}
\usepackage{enumitem}
\usepackage{bbm}
\usepackage{dsfont}
\usepackage{amssymb}

\usepackage{url}

\usepackage{subcaption}

\usepackage{bigstrut,multirow,rotating}
\usepackage{booktabs,chemformula}
\usepackage{placeins}

\graphicspath{ {figs/} }
\usepackage{algorithm,algorithmic}

\usepackage{balance}

\newcommand{\myparatight}[1]{\smallskip\noindent{\bf {#1}:}~}

\newenvironment{packeditemize}{\begin{list}{$\bullet$}{\setlength{\itemsep}{0.2pt}\addtolength{\labelwidth}{-4pt}\setlength{\leftmargin}{\labelwidth}\setlength{\listparindent}{\parindent}\setlength{\parsep}{1pt}\setlength{\topsep}{0pt}}}{\end{list}}

\begin{document}
	
\copyrightyear{2018} 
\acmYear{2018} 
\setcopyright{acmcopyright}
\acmConference[ACSAC '18]{2018 Annual Computer Security Applications Conference}{December 3--7, 2018}{San Juan, PR, USA}
\acmPrice{15.00}
\acmDOI{10.1145/3274694.3274706}
\acmISBN{978-1-4503-6569-7/18/12}	
	
\title{Poisoning Attacks to Graph-Based Recommender Systems}

\author{Minghong Fang}
\affiliation{%
  \institution{Iowa State University}
}
\email{myfang@iastate.edu}

\author{Guolei Yang}
\affiliation{%
	\institution{Facebook, Inc.}
}
\email{glyang@fb.com}

\author{Neil Zhenqiang Gong}
\affiliation{%
	\institution{Iowa State University}
}
\email{neilgong@iastate.edu}

\author{Jia Liu}
\affiliation{%
	\institution{Iowa State University}
}
\email{jialiu@iastate.edu}

\begin{abstract}
Recommender system is an important component of many web services to help users locate items that match their interests. Several studies showed that recommender systems are vulnerable to \emph{poisoning attacks}, in which an attacker injects fake data to a recommender system such that the system makes recommendations as the attacker desires. However, these poisoning attacks are either agnostic to recommendation algorithms or optimized to recommender systems (e.g., association-rule-based or matrix-factorization-based recommender systems) that are not graph-based. Like association-rule-based and matrix-factorization-based recommender systems, graph-based recommender system is also deployed in practice, e.g., eBay, Huawei App Store (a big app store in China). However, how to design optimized poisoning attacks for graph-based recommender systems is still an open problem. 

In this work, we perform a systematic study on poisoning attacks to graph-based recommender systems. We consider an attacker's goal is to promote a target item to be recommended to as many users as possible. To achieve this goal, our attacks inject fake users with carefully crafted rating scores to the recommender system. Due to limited resources and to avoid detection, we assume the number of fake users that can be injected into the system is bounded. The key challenge is how to assign rating scores to the fake users such that the target item is recommended to as many normal users as possible. To address the challenge, we formulate the poisoning attacks as an optimization problem, solving which determines the rating scores for the fake users. We also propose techniques to solve the optimization problem. We evaluate our attacks and compare them with existing attacks under white-box (recommendation algorithm and its parameters are known), gray-box (recommendation algorithm is known but its parameters are unknown), and black-box (recommendation algorithm is unknown) settings using two real-world datasets. Our results show that our attack is effective and outperforms existing attacks for graph-based recommender systems. For instance, when 1\% of users are injected fake users, our attack can make a target item recommended to 580 times more normal users in certain scenarios. 
  
\end{abstract}

%
%



\maketitle

\section{Introduction} \label{sec:intro}

In the era of big data, a fundamental challenge is to locate the data that are relevant to a particular user. Recommender systems aim to address this challenge: given a user's historical behavior and social data, a recommender system finds the data that match the user's preference. Indeed, recommender systems are widely deployed by web services (e.g., YouTube, Amazon, and Google News)  to recommend users relevant items such as products, videos, and news. 
In particular, \emph{collaborative filtering based recommender systems}, which analyze the correlations between users' historical behavior data for making recommendations, are widely deployed due to their effectiveness and generality. Depending on the techniques used to capture the correlations between users' behavior data, collaborative filtering based recommender systems can further include \emph{matrix-factorization-based}~\cite{MFRec09}, \emph{association-rule-based}~\cite{mobasher2000automatic,davidson2010youtube}, and \emph{graph-based}~\cite{fouss2007random} recommender systems. For instance, matrix-factorization-based recommender systems are deployed by Netflix to recommend movies, association-rule-based recommender systems are deployed by YouTube to recommend videos~\cite{davidson2010youtube}, and graph-based recommender systems are deployed by eBay~\cite{eBay2013a,eBay2013b} and Huawei App Store (a big app store in China)~\cite{He15,Guo17}. 


It is commonly believed that recommender systems recommend users items that match their personal interests. However, several studies~\cite{profileinjectionattacksMahony04,lam2004shilling,mobasher2007toward,poisoningattackRecSys16,yang2017fake} have demonstrated that recommender systems are vulnerable to \emph{poisoning attacks}, which inject fake data to a recommender system such that the recommender system makes recommendations as an attacker desires. For instance, an attacker can inject fake users with carefully crafted fake rating scores to a recommender system such that a target item is recommended to as many users as possible. Conventional poisoning attacks~\cite{profileinjectionattacksMahony04,lam2004shilling,mobasher2007toward} (also known as \emph{shilling attacks}) are agnostic to recommendation algorithms, i.e., they are not optimized to a certain type of recommender systems. Therefore, such attacks often achieve suboptimal performance when the recommendation algorithm is known. To address this limitation, recent studies~\cite{poisoningattackRecSys16,yang2017fake} proposed poisoning attacks that were optimized for a particular type of recommender systems. For instance,
 Li et al.~\cite{poisoningattackRecSys16} proposed poisoning attacks optimized for matrix-factorization-based recommender systems, while 
 Yang et al.~\cite{yang2017fake} proposed poisoning attacks optimized for association-rule-based recommender systems. However,  how to design optimized poisoning attacks to graph-based recommender systems is still an open problem.

In this work, we aim to design poisoning attacks for graph-based recommender systems~\cite{fouss2007random,eBay2013a,eBay2013b,He15,Guo17}. A graph-based recommender system uses a \emph{user preference graph} to represent users' rating scores to items. 
In the graph, a node is a user or an item, an edge between a user and an item means that the user rated the item, and the edge weight is the corresponding rating score. To make recommendations to a user, the recommender system performs a \emph{random walk} in the user preference graph, where the random walk starts from the user and jumps back to the user with a certain probability (called \emph{restart probability}) in each step. After the random walk converges, each item has a stationary probability that essentially characterizes the closeness between the item and the user. Finally, the system recommends the items that have the largest stationary probabilities to the user. 

In our poisoning attacks, an attacker's goal is to \emph{promote} a target item, i.e., making a graph-based recommender system  recommend the target item to as many users as possible.  
Like most existing poisoning attacks to recommender systems~\cite{profileinjectionattacksMahony04,lam2004shilling,mobasher2007toward,poisoningattackRecSys16}, our attacks inject fake users with carefully crafted rating scores to the target recommender system to achieve the attack goal. Due to limited resources and to avoid detection, we assume an attacker can inject $m$ fake users at most and each fake user rates $n$ items at most. For convenience, we call the items, which a fake user rates, the user's \emph{filler items}. The key challenge is to determine the filler items and their rating scores for each fake user. 
To address the challenge, we formulate poisoning attacks to graph-based recommender systems as an optimization problem,  whose {objective function} is the \emph{hit ratio} of the target item (i.e., the fraction of normal users whose recommended items include the target item) and whose {constraints} are that at most $m$ fake users with at most $n$ filler items can be injected. Solving this optimization problem produces $m$ fake users that maximize the hit ratio of the target item. 

However, this optimization problem is computationally intractable because  1) the hit ratio is related to the fake users' rating scores in a very complex way, and 2) the rating scores are integer variables. To address the computational challenge, we propose several techniques to solve the optimization problem approximately. Specifically, we approximate the hit ratio using the items' stationary probabilities, which are used to make recommendations in graph-based recommender systems. Moreover, 
we relax the fake users' rating scores as continuous variables, solve the optimization problem, and then generate filler items and their integer rating scores based on  the continuous variables. Finally, we propose a projected gradient descent based method to solve the optimization problem with an approximate hit ratio and relaxed continuous variables.

We evaluate our poisoning attacks and compare them with several existing attacks using two real-world datasets. First, we evaluate the attacks under the \emph{white-box setting}, i.e., the graph-based recommendation algorithm and its parameter (i.e., restart probability) are known to the attacker. We find that our attack can effectively enhance the hit ratio of a target item. For instance, when the  system recommends 10 items to each user and the number of injected fake users is 1\% of the number of normal users, our attack  could improve  the hit ratio of an unpopular target item by around 580 times. Moreover, our attack is significantly more effective than existing attacks for graph-based recommender systems. For instance, compared to the poisoning attack proposed by Yang et al.~\cite{yang2017fake}, our attack can improve the hit ratio from 0.0\% to 0.4\% for an unpopular target item. The reason is that existing attacks are not optimized for graph-based recommender systems.  Second, we evaluate the attacks under \emph{gray-box setting} (the graph-based recommendation algorithm is known but its parameter is unknown) and \emph{black-box setting} (the recommendation algorithm is unknown). We find that, in the gray-box setting, even if the attacker does not know the restart probability, our attack can still substantially improve the hit ratios of target items. In the black-box setting, we assume an attacker generates fake users based on a graph-based recommender system, while the target recommender system is based on matrix factorization. Our results show that our attacks can also transfer to matrix factorization based recommender systems.

We also study detecting fake users via supervised machine learning techniques and their impact on the effectiveness of poisoning attacks. Intuitively, the rating scores of fake users are generated in specific ways, and thus it could be possible to distinguish between normal users and fake users using their rating scores. Specifically, we extract features from a user's rating scores and learn a binary classifier using a training dataset that includes both normal users and fake users. The binary classifier is then used to predict a user to be normal or fake. We find that a small fraction of normal users are falsely predicted to be fake, while a large fraction (20\%$\sim$50\%) of fake users are falsely predicted to be normal. The service provider could deploy such a detector to predict fake users and exclude the predicted fake users from the recommender system. We evaluate our poisoning attacks and existing attacks under such scenario. We find that the poisoning attacks are still effective when such a detector is deployed, and our attack is still more effective than existing attacks. The reason is that a large fraction of fake users are not detected.

In summary, our contributions are as follows:
\begin{packeditemize}
\vspace{2mm}
\item We provide the first systematic study on poisoning attacks to graph-based recommender systems. We formulate poisoning attacks as an optimization problem and propose techniques to solve the optimization problem approximately.  

\vspace{2mm}
\item We extensively evaluate our attacks and compare them with existing attacks using two real-world datasets.  

\vspace{2mm}
\item We study detecting fake users using their rating scores and evaluate the effectiveness of poisoning attacks when such a detector is deployed.

\end{packeditemize}

\section{Background and Related Work} \label{sec:related}

\subsection{Collaborative Filtering}
Collaborative filtering based recommender systems have been widely deployed in various web services such as Amazon, YouTube, Netflix, and Google Play. Suppose we are given a \emph{user-item rating score matrix}, where the entry $r_{ui}$ is the rating score that user $u$ gave to item $i$, e.g., a product on Amazon, a video on YouTube, a movie on Netflix, and an app on Google Play. For instance, a rating score $r_{ui}$ can be 0, 1, 2, 3, 4, or 5, where $r_{ui}$=0 indicates that $u$ did not rate the item $i$, 1 means the most negative rating score, and 5 means the most positive rating score. The goal of collaborative filtering is to recommend \emph{each} user in the user-item rating score matrix $N$ items that the user did not rate before but the user may have interests in, via analyzing the rating score matrix. Depending on the techniques that are used to analyze the rating score matrix, collaborative filtering can be roughly classified to 4 categories, i.e., \emph{neighborhood-based}, \emph{association-rule-based}, \emph{matrix-factorization-based}, and \emph{graph-based}. 

\myparatight{Neighborhood-based, association-rule-based, and matrix-fac-torization-based recommender systems}
Neighborhood-based recommender systems~\cite{sarwar2001item} find neighbors of a user or neighbors of an item in order to recommend items to a user. For instance, to recommend a user items, the methods  can first find the nearest-neighbors of the user, predict the user's rating scores to items based on the rating scores of the nearest neighbors, and recommend the $N$ items that have the highest predicted rating scores to the user. 
Association-rule-based recommender systems~\cite{mobasher2000automatic,davidson2010youtube} aim to identify frequent co-occurrence between items in user reviews. For instance, if many users give high rating scores to both item $A$ and item $B$, then there is a certain association between the two items. For a user who gave a high rating score to item $A$, item $B$ is recommended to the user. 
Matrix-factorization-based recommender systems~\cite{MFRec09} assume that the user-item rating score matrix can be explained by a small number of latent factors. Based on the assumption, they use a low-rank matrix to approximate the user-item rating score matrix. The low-rank matrix predicts missing values in the user-item rating score matrix, i.e., for each user, the low-rank matrix predicts rating scores to all items that the user did not rate before; and the $N$ items that have the highest predicted rating scores are recommended to the user.

\begin{figure}[!t]
	\centering
	\includegraphics[scale = 0.55]{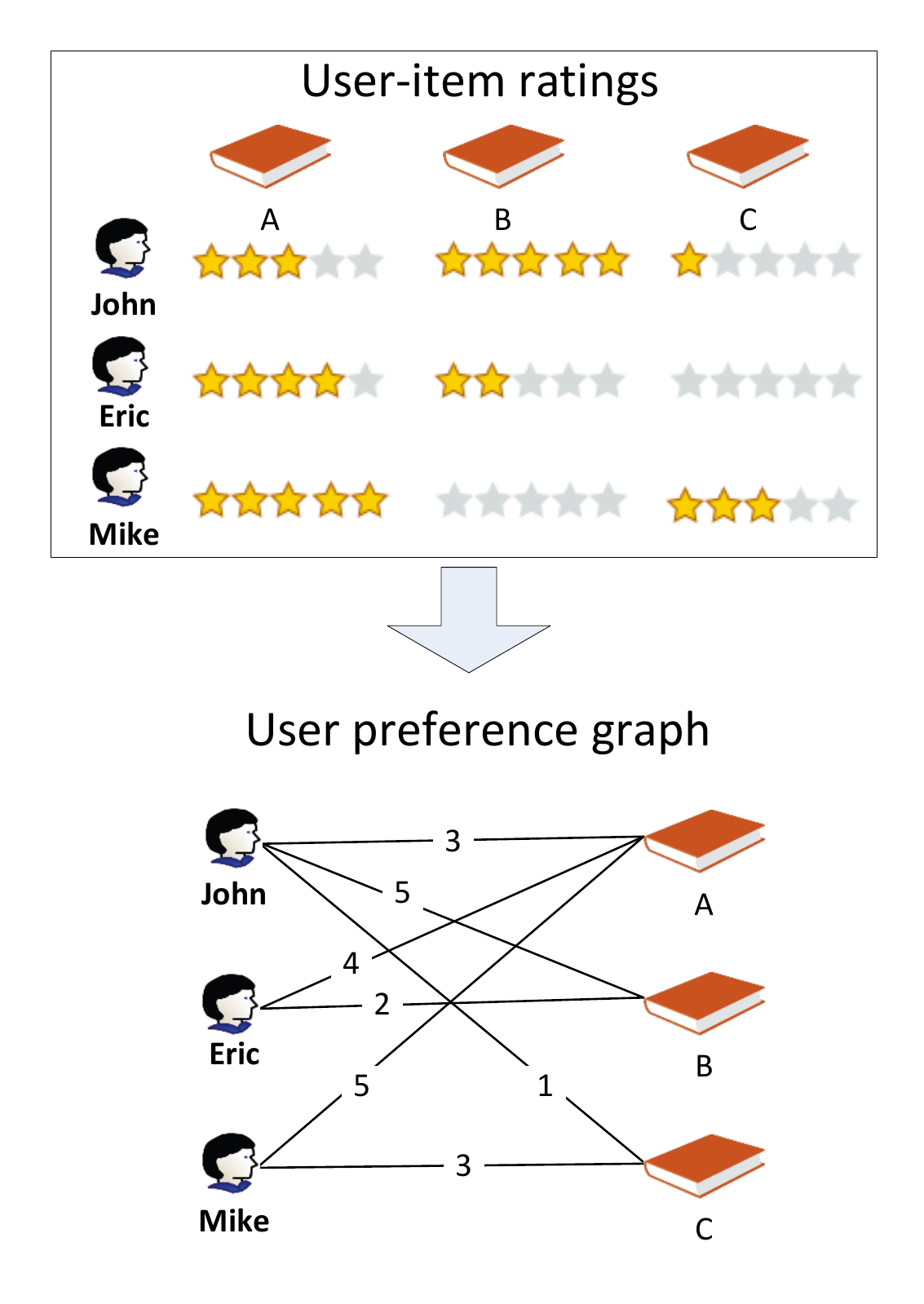}
	\caption{An illustration of user preference graph}\label{graph-based}
\end{figure}

\myparatight{Graph-based recommender systems} In this work, we focus on graph-based recommender systems~\cite{fouss2007random}. Graph-based recommender systems were deployed by several popular web services such as eBay~\cite{eBay2013a,eBay2013b} and Huawei App Store~\cite{He15,Guo17} in China. 
 The key idea of graph-based recommender system is to model users' preference for items as a weighted bipartite graph $G =  ( U, I, E )$, namely \emph{user preference graph}. The two sets of vertex $U$ and $I$ represent the user set and the item set, respectively;  an edge $(u,i)$ between a user $u \in U$ and an item $i \in I$ represents that the user rated the item; and the  weight of an edge $(u,i)$ is the rating score that the user gave to the item.  
Figure~\ref{graph-based} illustrates a user preference graph with an example of 3 users and 3 items. 

To generate the top-$N$ recommendation list for a user, the recommender system performs a \emph{random walk} in the graph, where the random walk starts from the user and jumps back to the user with a probability $\alpha$ in each step, where $\alpha$ is called \emph{restart probability}. The stationary probability distribution of the random walk is used to rank items and make recommendations. We denote by $p_u$ the stationary probability distribution of the random walk that starts from the user $u$. Then, the stationary probability distribution is a solution of the following linear system:
\begin{align}
\label{RWR}
p_u = (1 - \alpha ) \cdot Q \cdot p_u + \alpha  \cdot e_u,
\end{align}
where $e_u$ is a unit vector whose $u$th entry is 1 and all other entries are 0, and the matrix $Q$ is called \emph{transition matrix}, which is defined as follows:
\begin{align}
\label{matrixQ}
{Q_{xy}} = \begin{cases}
\frac{{{r_{xy}}}}{{\sum\limits_{z \in {\Gamma_x}} {{r_{xz}}} }} & \mbox{if $(x, y) \in E$}\\
0 & \mbox{otherwise,}
\end{cases}
\end{align}
where $\Gamma_x$ is the set of neighbors of node $x$. More specifically, for a user node $x$, $\Gamma_x$ is the set of items that were rated by $x$; for an item node $x$, $\Gamma_x$ is the set of users that rated $x$.
To solve $p_u$, we start from a random probability distribution and then iteratively update $p_u$ as $p_u^{(t+1)} = (1 - \alpha ) \cdot Q \cdot p_u^{(t)} + \alpha  \cdot e_u$ until convergence. Then, we rank the items that were not rated by the user $u$ with respect to their stationary probabilities. The top-$N$ items with the largest stationary probabilities are recommended to the user $u$.


\subsection{Attacks to Recommender Systems}


\subsubsection{Security Attacks} These attacks aim to spoof a recommender system such that a target item is recommended to as many or few users as possible.  Specifically, \emph{poisoning attacks} (also known as \emph{shilling attacks})~\cite{profileinjectionattacksMahony04,lam2004shilling,mobasher2007toward} aim to inject fake users with fake rating scores to the system such that a bad recommender system is learnt from the user-item rating score matrix. \emph{Profile pollution attacks}~\cite{xing2013take} aim to pollute the rating behavior of normal users to manipulate the recommendations to them. By analogy to adversarial machine learning,  poisoning attacks are to manipulate recommender systems at ``training time", while profile pollution attacks are to manipulate recommender systems at ``testing time".

\myparatight{Poisoning attacks}   Poisoning attacks were first studied more than a decade ago~\cite{profileinjectionattacksMahony04,lam2004shilling,mobasher2007toward}. However, these attacks are heuristics-driven and are not optimized to a particular type of recommender systems. For instance, in \emph{random attacks}~\cite{lam2004shilling}, given the number of fake users an attacker can inject into the system, the attacker randomly selects some items for each fake user and then generates a rating score for each selected item from a normal distribution, whose mean and variance are calculated from the rating scores in the entire user-item rating score matrix. In \emph{average attacks}~\cite{lam2004shilling}, the attacker generates a rating score for a selected item from a normal distribution, whose mean and variance are computed from the rating scores to the selected item in the user-item rating score matrix. 

More recent poisoning attacks~\cite{poisoningattackRecSys16,yang2017fake} generate fake rating scores or behavior that are optimized to a particular type of recommender systems.  Specifically, Li et al.~\cite{poisoningattackRecSys16} proposed poisoning attacks to matrix-factorization-based recommender systems. Yang et al.~\cite{yang2017fake} proposed poisoning attacks (they called them \emph{fake co-visitation injection attacks}) to association-rule-based recommender systems, in which each user injects fake co-visitations between items instead of fake rating scores to items.  We aim to study optimized  poisoning attacks to graph-based recommender systems.

\myparatight{Profile pollution attacks} Xing et al.~\cite{xing2013take} proposed {profile pollution attacks} to recommender systems and other personalized services, e.g., web search. Their attacks aim to pollute a user's profile, e.g., browsing history, via cross-site request forgery (CSRF)~\cite{CSRF}. With a polluted user profile, the attacker can recommend arbitrary items to the user. 
They showed that  popular web services including YouTube, Amazon, and Google search are vulnerable to the attacks. However, the limitation of these attacks is that they rely on CSRF, which makes it hard to perform the attacks at a large scale.



\subsubsection{Privacy Attacks} Two attacks, i.e., \emph{item inference attacks} and \emph{attribute inference attacks}, were proposed to compromise user privacy in recommender systems.

\myparatight{Item inference attacks} Calandrino et al.~\cite{CalandrinoOakland11} proposed privacy attacks to infer the items that a target user has rated before, e.g., such items could be products that the target user purchased on Amazon, music the target user liked on Last.fm, and books the target user read on LibraryThing. The key intuition of their attacks is that a collaborative filtering recommender system makes recommendations based on users' past behavior. Therefore, the recommendations made by a recommender system include information about users' past behavior. Via tracking and analyzing the publicly available recommendations over time, an attacker could infer a target user's past behavior, e.g., the items the user rated.


\myparatight{Attribute inference attacks} A user's rating behavior (e.g., rating scores to items, page likes on Facebook) is essentially statistically correlated to the user's attributes (e.g., gender, political view, sexual orientation, interests, and location). Therefore, an attacker could infer a user's private attributes based on its rating behavior via machine learning techniques, which capture the statistical correlations between rating behavior and attributes. Such attacks are called \emph{attribute inference attacks}~\cite{GongAttriInferSEC16} and have been demonstrated to be feasible by multiple studies~\cite{weinsberg2012blurme,kosinski2013private,Gong14,GongAttriInferSEC16,AttriInfer,gongTOPS18}. In particular, given a set of users whose rating behavior and attributes are known to an attacker, the attacker learns a machine learning classifier which takes a user's rating behavior as an input and predicts the user's attributes. Then, the attacker applies this classifier to infer attributes of the users who did not disclose their attributes. A notable example of real-world attribute inference attacks is that Cambridge Analytica leveraged Facebook users' rating behavior (e.g., page likes) to infer users' attributes, based on which targeted advertisements are delivered to users~\cite{Cambridge}. Jia and Gong~\cite{Jia18} recently proposed a practical defense against attribute inference attacks via adversarial machine learning. The key idea is to add carefully crafted noise to a user's rating behavior data such that the attacker's classifier is very likely to make incorrect predictions.

\section{Problem Formulation} \label{sec:problem}

\subsection{Threat Model}

\myparatight{Attack goal} We consider an attacker's goal is to \emph{promote} a target item $t$ to as many users as possible. 
Suppose the system recommends $N$ items to each user. We denote by $h(t)$ the fraction of normal users whose top-$N$ recommendations include the target item after the attack. $h(t)$ is called \emph{hit ratio} of the target item $t$. The attacker's goal is to maximize the hit ratio. We note that an attacker could also \emph{demote} a target item, i.e., minimize the hit ratio of the target item. However, demotion is a special case of promotion~\cite{mobasher2007toward,yang2017fake}. Specifically, an attacker can promote other items such that the target item is demoted in recommendation lists. Therefore, we will focus on promotion attacks in this work.


\myparatight{Attack approach} The attacker uses data poisoning attacks to achieve the attack goal. In particular, the attacker injects some fake users to the system. Each fake user gives a high rating score to the target item and well-crafted rating scores to certain selected items, which we call \emph{filler items}. A key challenge for the attacker is to determine the filler items and their rating scores for each fake user. Since normal users often rate a small number of items, we assume the number of filler items for each fake user is at most $n$, to avoid being detected simply based on the number of rated items. 


\myparatight{Attacker's background knowledge and capability} We assume an attacker has the following background knowledge: 1) the recommendation algorithm used by the given recommender system; and 2) the user-item rating score matrix, which is usually publicly available and can be collected by the attacker. We note that the attacker could also collect a partial  user-item rating score matrix for a subset of users and subset of items, and design attacks based on the partial matrix. Our threat model is also known as \emph{white-box} setting. In our experiments, we will demonstrate that our attacks can also be transferred between recommender systems under the \emph{grey-box} setting (i.e., the attacker does not know the parameters of the recommendation algorithm) or the \emph{black-box} setting (i.e., the attacker does not know the recommendation algorithm). 

In practice, an attacker often has limited resources so the attacker can only inject a bounded number of fake users into the system, though the bounded number could still be large. For instance, an attacker could leverage compromised machines to register and maintain fake users. Detecting such fake users is also known as \emph{Sybil detection}, and many methods (e.g.,~\cite{Wang13Clickstream,EVILCOHORTSecurity15,sybilbelief}) have been developed to detect fake users. For instance, the service provider could analyze the IP addresses of the users to detect fake ones. To avoid such IP-based detection, 
an attacker often registers a small number of fake users on a compromised machine. Indeed, Thomas et al.~\cite{SybilUnderground13} found that a half of compromised machines under an attacker's control maintain less than 10 fake users in online social networks. 
More formally, we assume the attacker can inject $m$ fake users into the recommender system.

\subsection{Attacks as an Optimization Problem}


We formulate poisoning attacks as an optimization problem, solving which maximizes the hit ratio of the target item. 
Let $r_v$ be the rating score vector of a fake user $v$, where $r_{vi}$ is the rating score that the fake user $v$ gives to the item $i$. We consider a rating score is in the set of integers $\{0, 1, \cdots, r_{max}\}$, where $r_{max}$ is the maximum rating score. For instance, in many recommender systems, $r_{max}=5$. A rating score of 0 means that the user did not rate the corresponding item. Essentially, we aim to find the rating score vector for each fake user that maximizes the hit ratio of the target item. Specifically, we find the rating score vectors via solving the following optimization problem:
\begin{align}
\label{objectiveFrame}
\text{max }  &  h(t)    \\
\text{subject to } & |r_v|_0 \leq n + 1, \forall v\in \{v_1, v_2, \cdots, v_m\}  \nonumber \\
& r_{vi} \in \{0,1,\cdots,r_{max}\},  \forall v\in \{v_1, v_2, \cdots, v_m\}, \nonumber
\end{align}
where $\{v_1, v_2, \cdots, v_m\}$ is the set of $m$ fake users, $|r_v|_0$ is the number of non-zero entries in the rating score vector $r_v$, and $n$ is the maximum number of filler items (the filler items do not include the target item). The hit ratio $h(t)$, which is the fraction of normal users whose top-$N$ recommended items include the target item $t$, is computed by a  recommender system on the entire user-item rating score matrix that includes the $m$ fake users. We note that our formulation in Equation~\ref{objectiveFrame} is applicable to data poisoning attacks to any recommender system. In this work, we will focus on graph-based recommender systems.

\section{Our Poisoning Attacks} \label{AttackModel}
\subsection{Overview}
A solution to the optimization problem in Equation~\ref{objectiveFrame} is a data poisoning attack. 
However, finding the exact optimal solution to the optimization problem in Equation~\ref{objectiveFrame} is computationally intractable (i.e., NP-hard) because 1) the objective function $h(t)$ is related to the rating score variables $r_v$ ($v\in \{v_1,v_2,\cdots, v_m\}$) in a very complex way, and 2) the variables are integer variables. Therefore, we propose techniques to find approximate solutions to the optimization problem. 


Specifically, to address the computational challenge, we propose several approximation techniques. First, instead of  optimizing the rating scores for the $m$ fake users simultaneously, we optimize their rating scores one by one. In particular, given the normal users and fake users we have added so far, we find the rating scores for the next fake user to optimize the hit ratio of the target item. Second, we approximate the hit ratio $h(t)$ in the objective function using some function that is easier to optimize. Specifically, since graph-based recommender systems leverage the stationary probabilities of items to make recommendations, our approximate objective function roughly requires that the stationary probabilities of the target item are high for many users. Third, we relax the rating scores to be continuous variables in the range [0, $r_{max}$] and then transform them to integer rating scores after solving the  optimization problem. We propose a projected gradient descent based method to solve the optimization problem with the approximate objective function and relaxed continuous variables.

\subsection{Approximating the Optimization Problem}
Suppose $t$ is the target item that the attacker aims to promote. We add fake users to the recommender system one by one. Assume $G =  ( U, I, E )$ is the current user preference graph which includes rating scores for both normal users and fake users added so far.  
$S$ is the set of normal users who have not rated the target item $t$. We denote the set of top-$N$ recommended items for a user $u$ as ${L_u}$.

\myparatight{Relaxing rating scores to be continuous variables} We add a fake user $v$ to the user preference graph $G$, where $w_{vi}$ is the rating score that the fake user gives to item $i$. We model $w_{vi}$ as the weight of the edge $(v,i)$. For simplicity, we denote by $w_v$ the vector of weights of edges that connect the fake user $v$ and all items. 
Our goal is to find the edge weights $w_v$ that optimize the hit ratio of the target item.  Since rating scores are integers, $w_v$ are integer variables whose values could be $0,1,\cdots, r_{max}$. However, such integer variables make the optimization problem intractable. Therefore, we  relax the variables  as continuous variables whose values are in the range [0, $r_{max}$], solve the optimization problem using the continuous variables, and transform them to integer rating scores. Note that $w_{vi}$ is different from $r_{vi}$. Specifically, $w_{vi}$ is a continuous variable we use to model a rating score, while $r_{vi}$ is the final integer rating score that user $v$ gives to item $i$. 

\myparatight{Approximating the hit ratio} Since the hit ratio is related to the edge weights $w_v$ in a very complex way, which makes the optimization problem intractable, we approximate the hit ratio using the stationary probabilities of random walks, which are used to generate the top-$N$ recommended items in graph-based recommender systems. 
 In the user preference graph with the new fake user $v$, to make recommendations for a normal user $u$, we first perform a random walk from $u$ and compute its stationary probability distribution $p_u$, where $p_{ui}$ is the stationary probability for item $i$. Specifically, the stationary probability distribution $p_u$ is computed according to Equation~\ref{RWR}, where the transition matrix $Q$ is a function of the edge weights $w_v$. The recommendation list $L_u$ consists of the $N$ items that 1) $u$ has not rated yet and 2) have the largest stationary probabilities. The target item $t$ hits the user $u$ if $t$ is among the recommendation list $L_u$, i.e., if $p_{ut}> p_{ui}$ for a certain item $i$ in the recommendation list $L_u$, otherwise the target item does not hit the user $u$.

{\bf 1) Loss function for one user.} To approximate the hit ratio, we leverage a \emph{loss function} $l_u$ over the stationary probability distribution for each user $u$. We aim to design a loss function that satisfies two goals: 1) for each item $i\in L_u$, if $p_{ui} < p_{ut}$ (i.e., the target item ranks before the item $i$), then the loss for item $i$ is smaller,  and 2) the loss is smaller if the target item ranks higher in the recommendation list $L_u$. To achieve these goals, we adopt the following loss function:
\begin{align}
l_u=\sum_{i \in L_u} g(p_{ui} - p_{ut}),
\end{align} 
where $g(x) = \frac{1}{{1 + \exp (-{x/b})}}$ is called the Wilcoxon-Mann-Whitney loss function~\cite{backstrom2011supervised} and $b$ is a parameter called \emph{width}. In the machine learning community, the Wilcoxon-Mann-Whitney loss function is known to optimize the ranking performance~\cite{backstrom2011supervised}, i.e., the loss is smaller when the target item ranks higher in the recommendation list in our case. 

{\bf 2) Loss function for all normal users.} Our goal is to recommend the target item to as many normal users as possible. Therefore, we sum the loss of all normal users as follows:
\begin{align}
l=\sum_{u \in S} l_u,
\end{align}
where $S$ is the set of normal users who have not rated the target item yet.

{\bf 3) Approximate optimization problem.} Recall that, in our threat model, each fake user rates at most $n$ items to avoid detection, which essentially constrains the values of $w_v$. Considering this constraint, we propose to solve the following optimization problem:
\begin{align}
\min F(w_v) &= {\left\| w_v \right\|_2^2} + \lambda \cdot l \nonumber\\
\label{appproblem}
\text{subject to } &w_{vi} \in [0, r_{max}],
\end{align}
where $\left\| w_v \right\|_2^2$ regularizes $w_v$ and is used to model the constraint that each fake user can rate a small number of items, while $\lambda$ balances the regularization term and the loss function.

\begin{algorithm}[!t]
	\caption{\textit{Our Poisoning Attacks}}\label{attack_algo}
	\begin{algorithmic}[1]
		\renewcommand{\algorithmicrequire}{\textbf{Input:}}
		\renewcommand{\algorithmicensure}{\textbf{Output:}}
		\REQUIRE  Rating matrix $R$, parameters $t, m, n, \lambda, b$.
		\ENSURE  $m$ fake users $v_1, v_2, \cdots, v_m$.
		\STATE //Add fake users one by one.
		\FOR {$ v = v_1, v_2, \cdots, v_m$}
			\STATE Solve the optimization problem in Equation \ref{appproblem} with the current rating matrix $R$ to get $w_v$.			
			\STATE //Assign the maximum rating score to the target item.
			\STATE $r_{vt} =  r_{max}$.
			\STATE //Find the filler items
			\STATE The $n$ items with the largest weights are filler items.
			\STATE //Generate rating scores for the filler items.
			\STATE  $r_{vj} \sim \mathcal{N}(\mu_{j},\,\sigma_{j}^{2})$, for each filler item $j$. 
			\label{normaldistribution}
			\STATE //Inject the fake user with rating scores $r_v$ to the system.
			\STATE $R\leftarrow R \cup r_v$.  
		\ENDFOR
		\RETURN $r_{v_1}, r_{v_2}, \cdots, r_{v_m}$. 
	\end{algorithmic} 
\end{algorithm}

\subsection{Solving the Optimization Problem}
We solve the optimization problem in Equation~\ref{appproblem} using \emph{projected gradient descent}. Specifically, in each iteration, we compute the gradient of $F(w_v)$ with respect to $w_v$, move $w_v$ a small step towards the inverse direction of the gradient, and project each $w_{vi}$ back to the range $[0,r_{max}]$. We can compute the gradient of $F(w_v)$ as follows: 
\begin{align}
\label{rwrOPtMIN}
\begin{split}
\frac{{\partial F(w_v)}}{{\partial w_v}} &= 2w_v + \lambda \sum_{u \in S} {\sum_{i \in L_u} {\frac{{\partial g(p_{ui} - p_{ut})}}{{\partial w_v}}}} \\
&= 2w_v + \lambda \sum_{u \in S} {\sum_{i \in L_u} {\frac{{\partial g({\delta _{it}})}}{{\partial {\delta _{it}}}}} } (\frac{{\partial p_{ui}}}{{\partial w_v}} - \frac{{\partial {p_{ut}}}}{{\partial w_v}}),
\end{split}
\end{align}
where ${\delta _{it}} = p_{ui} - p_{ut}$. 

The key challenge of computing the gradient is to compute the gradient $\frac{{\partial {p_u}}}{{\partial w_v}}$ for each normal user $u$.
 From Equation~\ref{RWR}, we have:
\begin{align}
\label{p_to_w}
\frac{{\partial {p_u}}}{{\partial w_v}} = (1 - \alpha )\frac{{\partial Q}}{{\partial w_v}}{p_u} + (1 - \alpha )Q\frac{{\partial {p_u}}}{{\partial w_v}}.
\end{align}

Furthermore, according to Equation~\ref{matrixQ}, we have: 
\begin{align}
\label{Q_to_w}
\frac{{\partial {Q_{xy}}}}{{\partial w_v}} = 
\begin{cases}
\frac{{\frac{{\partial {w_{xy}}}}{{\partial w_v}}\sum\nolimits_j {{w_{xj}} - {w_{xy}}\sum\nolimits_j {\frac{{\partial {w_{xj}}}}{{\partial w_v}}} } }}{{{{(\sum\nolimits_j {{w_{xj}}} )}^2}}}, \text{ if } (x, y) \in E \\
0, \text{ otherwise,}
\end{cases}
\end{align}
where $w_{xy}$ is the discrete rating score that user $x$ gave to the item $y$ when $x$ is not the new fake user, and $w_{xy}$ is the continuous edge weight to be optimized when $x$ is the new fake user. 
Therefore, Equation~\ref{p_to_w} is a linear system of equations with respect to $\frac{{\partial {p_u}}}{{\partial w_v}}$. We iteratively solve the linear system to obtain $\frac{{\partial {p_u}}}{{\partial w_v}}$. After solving $\frac{{\partial {p_u}}}{{\partial w_v}}$, we can compute the gradient 
$\frac{{\partial F(w_v)}}{{\partial w_v}}$.



\subsection{Generating Rating Scores}
After solving the weights $w_v$, we generate rating scores for the fake user $v$. First, we assume the fake user gives the maximum rating score to the target item. Second, we rank the items according to the weights $w_{vi}$ and select the $n$ items with the highest weights as the filler items. The fake user only generates rating scores for the filler items. Third, for each filler item, we sample a number  from a normal distribution that is fitted to the rating scores that all normal users gave to the item, and then discretize the number to an integer rating score. We only use the weights to select filler items instead of assigning their rating scores, because the weights are approximate values. We generate rating scores for the filler items from such a normal distribution so that the fake user is  likely to be similar to more normal users, which makes it more likely to recommend the target item to more normal users. 

Algorithm~\ref{attack_algo} summarizes our poisoning attacks. We generate fake users one by one. For each fake user, we use projected gradient descent to solve the optimization problem in Equation~\ref{appproblem} with the current rating score matrix (i.e., the current user preference graph). After solving the weights $w_v$, we generate rating scores. Specifically, $\mathcal{N}(\mu_{j},\,\sigma_{j}^{2})$ at Line~\ref{normaldistribution} is the normal distribution with mean $\mu_{j}$ and variance $\sigma_{j}^{2}$ that are fitted using the rating scores that normal users gave to the item $j$.

\section{Experiments} \label{sec:exp}

\begin{table}[!t]
\centering
\caption{Dataset statistics.}\label{datasets}
\begin{tabular}{ccccc} 
\hline
Dataset &\#Users  &\#Items  &\#Ratings  &Sparsity\\ \hline  
Movie &943 &1,682 &100,000 &93.67\% \\     
Video &5,073 &10,843 &48,843 &99.91\%\\ \hline
\end{tabular}
\end{table}

\subsection{Experimental Setup}

\subsubsection{Datasets}
We perform experiments using two real-world datasets, which are widely used for evaluating recommender systems in the data mining community. The first dataset is \textbf{MovieLens 100K (Movie)}~\cite{movielensURL}. This dataset consists of 943 users, 1,682 movies, and 100,000 ratings. 
The second dataset is \textbf{Amazon Instant Video (Video)}~\cite{amazonURL}, which includes 5,073 users, 10,843 items, and 48,843 ratings. 
We define the \textit{sparsity} of a dataset as follows:
\begin{align}
Sparsity = 1 - \frac{\text{number of ratings}}{\text{number of users} \times \text{number of items}}.
\end{align}
As we will show, the attack performance is related to the sparsity of a recommender system.  Table~\ref{datasets} shows the dataset statistics. 


\subsubsection{Compared Attacks} We compare our poisoning attacks to several poisoning attacks. In all these attacks, an attacker injects $m$ fake users to the recommender system. Each fake user gives the maximum rating score to the target item and gives certain rating scores to $n$ selected items (called \emph{filler} items). Different attacks use different strategies to select the filler items and generate rating scores for them. 

\myparatight{Random attack~\cite{lam2004shilling}} In this attack, the attacker first fits a normal distribution for the rating scores in the entire user-item rating score matrix.  For each fake user, the attacker selects $n$ items as the filler items uniformly at random. Then, for each filler item, the attacker samples a number from the normal distribution and discretizes it to be a rating score. 


\myparatight{Average attack~\cite{lam2004shilling}} In this attack, the attacker fits a normal distribution for the rating scores of each item. Like the random attack, average attack also samples $n$ items as filler items uniformly at random. However, for each filler item, the attacker generates a rating score from the normal distribution fitted for the item. The intuition is that generating rating scores around the average rating scores of filler items could enable the fake users to be more similar to normal users, and thus have a larger effect on the recommendations.


\myparatight{Bandwagon attack~\cite{mobasher2007toward}} This attack considers item popularity when selecting filler items. We implement a variant of bandwagon attack as follows: for each fake user, the attacker selects $n\times 10\%$ items whose average rating scores are high (e.g., 5 in our experiments) and selects $n\times 90\%$ items uniformly at random as filler items. For each filler item, the attacker generates a rating score from the normal distribution fitted for the entire user-item rating score matrix (like the random attack). The intuition is that the attacker aims to recommend the target item to users who rated the popular items. 


\myparatight{Co-visitation attack~\cite{yang2017fake}} This attack was designed for association-rule-based recommender systems. We note that in the original attack, the attacker does not necessarily need to register fake users, because some association-rule-based recommender systems consider visitations from any visitors to make recommendations. In our work, we focus on recommender systems using rating scores and only registered users can provide rating scores. Therefore, the attacker injects fake users to the system. Moreover, if a user rates both items $i$ and $j$, then we say $i$ and $j$ are co-visited by the user. Therefore, the attack technique developed by Yang et al.~\cite{yang2017fake} essentially finds the filler items for each fake user. For each filler item of each fake user, we generate a rating score from the normal distribution fitted for the item (like the average attack). 


\subsubsection{Target Items (Random Target Items vs. Unpopular Target Items)} We consider two types of target items. First, an attacker aims to promote a \emph{random} target item.  Specifically, in our experiments, we sample an item uniformly at random and treat it as the target item. Second, an attacker could also promote an unpopular item (e.g., a new item that belongs to the attacker). To simulate this attacker, we sample an item that has 5 ratings at most uniformly at random and treat it as the target item. 

\subsubsection{Evaluation Metric (HR@N)} We use the hit ratio (HR@N) as our evaluation metric. Suppose the recommender system recommends $N$ items for each user. Given a target item, HR@N is the fraction of normal users whose $N$ recommended items include the target item. For both random target items and unpopular target items, we compute the hit ratio averaged over 10 target items. 


\subsubsection{Parameter Setting} Without otherwise mentioned, we use the following \emph{default parameter setting}: the restart probability $\alpha$ in graph-based recommender systems is set to be 0.3, $\lambda = 0.01, b = 0.01, N = 10$, and $n=10$. Moreover, the number of fake users (i.e., \emph{attack size}) is 3\% of the normal users in the recommender system. By default, we assume graph-based recommender system is used. 

\begin{table*}[ht]
	\centering
	\caption{HR@10 for different  attacks with different attack sizes. }\label{Data_poisoning_attacks_Results}
	\begin{tabular}{c|c|cccc|cccc}
		\hline
		\multirow{3}[6]{*}{Dataset} & \multirow{3}[6]{*}{Attack } & \multicolumn{8}{c}{Attack size} \bigstrut\\
		\cline{3-10}          &       & \multicolumn{4}{c|}{Random target items} & \multicolumn{4}{c}{Unpopular target items  } \bigstrut\\
		\cline{3-10}          &       & 0.5\% & 1\%   & 3\%   & 5\%   & 0.5\% & 1\%   & 3\%   & 5\% \bigstrut\\
		\hline
		\hline
		\multirow{6}[2]{*}{Movie} & None  & 0.0022 & 0.0022 & 0.0022 & 0.0022 & 0     & 0     & 0     & 0 \bigstrut[t]\\
		& Random & 0.0028 & 0.0030 & 0.0038 & 0.0052 & 0     & 0     & 0     & 0 \\
		& Average & 0.0027 & 0.0030 & 0.0038 & 0.0049 & 0     & 0     & 0     & 0 \\
		& Bandwagon & 0.0027 & 0.0030 & 0.0037 & 0.0048 & 0     & 0     & 0     & 0 \\
		& Co-visitation & 0.0030 & 0.0030 & 0.0037 & 0.0050 & 0     & 0     & 0.0005 & 0.0027 \\
		& Our attack & \textbf{0.0040} & \textbf{0.0069} & \textbf{0.0134} & \textbf{0.0168} & \textbf{0.0005} & \textbf{0.0042} & \textbf{0.0104} & \textbf{0.0131} \bigstrut[b]\\
		\hline
		\multirow{6}[2]{*}{Video} & None  & 0.0019 & 0.0019 & 0.0019 & 0.0019 & 0.0001 & 0.0001 & 0.0001 & 0.0001 \bigstrut[t]\\
		& Random & 0.0181 & 0.0377 & 0.1456 & 0.2692 & 0.0137 & 0.0317 & 0.1323 & 0.2500 \\
		& Average & 0.0185 & 0.0397 & 0.1472 & 0.2775 & 0.0148 & 0.0323 & 0.1358 & 0.2554 \\
		& Bandwagon & 0.0171 & 0.0372 & 0.1443 & 0.2660 & 0.0130 & 0.0314 & 0.1305 & 0.2481 \\
		& Co-visitation & 0.0180 & 0.0378 & 0.1460 & 0.2688 & 0.0135 & 0.0313 & 0.1333 & 0.2579 \\
		& Our attack & \textbf{0.0323} & \textbf{0.0625} & \textbf{0.1828} & \textbf{0.2966} & \textbf{0.0285} & \textbf{0.0576} & \textbf{0.1727} & \textbf{0.2845} \bigstrut[b]\\
		\hline
		\hline
	\end{tabular}%
\end{table*}%

\begin{table}[ht]
	\centering
	\caption{HR@$N$ for different $N$.}\label{TOPN}
	\addtolength{\tabcolsep}{-2pt}

	\begin{tabular}{c|c|ccccc}
		\hline
		\multirow{2}[4]{*}{Dataset} & \multirow{2}[4]{*}{Attack} & \multicolumn{5}{c}{$N$} \bigstrut\\
		\cline{3-7}          &       & 1     & 5     & 10    & 15    & 20 \bigstrut\\
		\hline
		\hline
		\multirow{6}[2]{*}{Movie} & None  & 0     & 0.0001 & 0.0022 &  0.0060 & 0.0085 \bigstrut[t]\\
		& Random & 0     & 0.0004 & 0.0038 & 0.0076 & 0.0109 \\
		& Average & 0     & 0.0005 & 0.0038 & 0.0077 & 0.0112 \\
		& Bandwagon & 0     & 0.0004 & 0.0037 & 0.0076 & 0.0109 \\
		& Co-visitation & 0     & 0.0007 & 0.0040 & 0.0074 & 0.0108 \\
		& Our attack & \textbf{0.0024} & \textbf{0.0066} & \textbf{0.0134} & \textbf{0.0193} & \textbf{0.0243} \bigstrut[b]\\
		\hline
		\multirow{6}[2]{*}{Video} & None  & 0.0001 & 0.0008 & 0.0019 & 0.0036 & 0.0047 \bigstrut[t]\\
		& Random & 0.0461 & 0.0989 & 0.1456 & 0.1820 & 0.2130 \\
		& Average & 0.0476 & 0.1019 & 0.1472 & 0.1840 & 0.2144 \\
		& Bandwagon & 0.0454 & 0.0975 & 0.1443 & 0.1783 & 0.2090 \\
		& Co-visitation & 0.0479 & 0.1018 & 0.1463 & 0.1835 & 0.2131 \\
		& Our attack & \textbf{0.0665} & \textbf{0.1359} & \textbf{0.1828} & \textbf{0.2116} & \textbf{0.2314} \bigstrut[b]\\
		\hline
		\hline
	\end{tabular}%
\end{table}%

\subsection{Attacking Graph-based Systems}
We first consider the white-box setting, i.e., the graph-based recommender system and its restart probability are known to the attacker. 

\myparatight{Impact of attack size} Table~\ref{Data_poisoning_attacks_Results} shows the results for the compared poisoning attacks with different attack sizes. The attack size means that the number of fake users is a certain fraction of the normal users, e.g., 1\% attack size means that the number of fake users is 1\% of the number of normal users. The row in ``None'' means the hit ratios without any attacks. First, our attack can effectively promote target items. For instance, in the Video dataset, when injecting 1\% fake users, the hit ratio of a random target item increases by around 33 times, while the hit ratio of an unpopular target item increases by around 580 times. Second, our attack is significantly more effective than existing attacks. For instance, in the Movie dataset, when injecting 1\% fake users, our attack improves the hit ratio upon the best compared attack by 2.3 times for a random target item, while our attack improves the hit ratio from 0 to 0.0042 for an unpopular target item. The reason is that random attack, average attack, and bandwagon attack are agnostic to recommender systems, while the co-visitation attack was specifically designed for association-rule-based recommender systems. 

Third, the hit ratio gain is more significant for unpopular target items than random target items. For instance, our attack improves the hit ratio by 96 times and 1700 times for a random target item and an unpopular target item respectively, when injecting 3\% fake users into the Video dataset. Fourth, all attacks are more effective on the Video dataset than the Movie dataset. We speculate the reason is that Video is more sparse, and thus is easier to attack. More specifically, when the dataset is more sparse, it is easier to inject fake users that are similar to a large number of normal users.

\myparatight{Impact of the number of recommended items} Table~\ref{TOPN} shows the hit ratios for different attacks when the recommender system recommends different numbers (i.e., $N$) of items to users, where random target items are used and the attack size is fixed to be 3\%. First, we observe that our attack is effective and is more effective than the existing attacks for different values of $N$. Second, when $N$ is smaller, the hit ratio gains of our attack over existing attacks are more significant. For instance, when $N=20$ and $N=5$, our attack's hit ratios improve upon the best existing attacks by twice and by 9.5 times in the Movie dataset, respectively. This indicates that our attack ranks the target item higher in the recommendation lists than existing attacks. The reason is that the Wilcoxon-Mann-Whitney loss function~\cite{backstrom2011supervised} adopted by our attacks aims to optimize the ranking performance of the target item.


\begin{figure}[t]
	\centering
	\begin{subfigure}[b]{0.23\textwidth}
		\includegraphics[width=\textwidth]{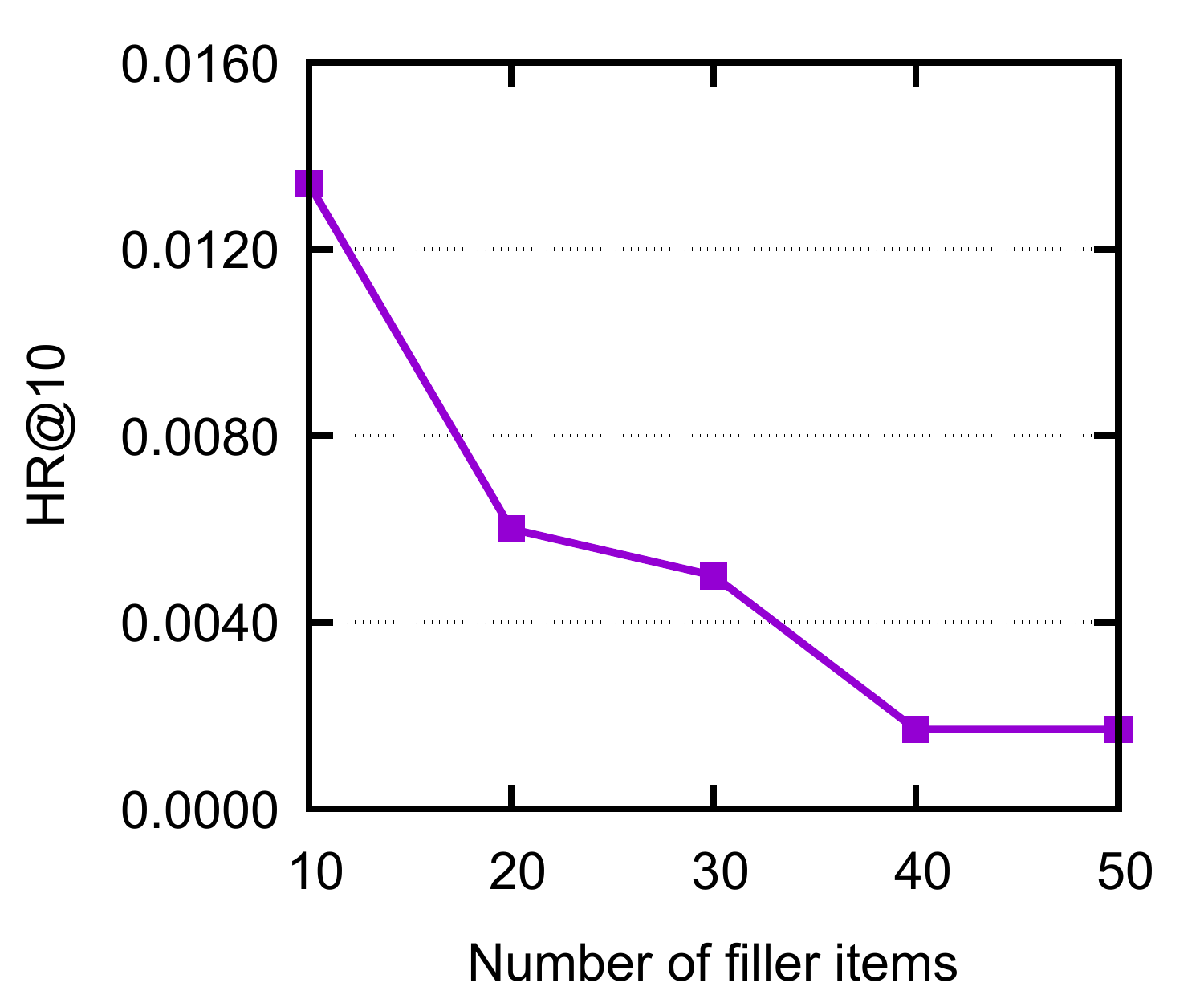}
		\caption{Movie}
	\end{subfigure}
	\begin{subfigure}[b]{0.23\textwidth}
		\includegraphics[width=\textwidth]{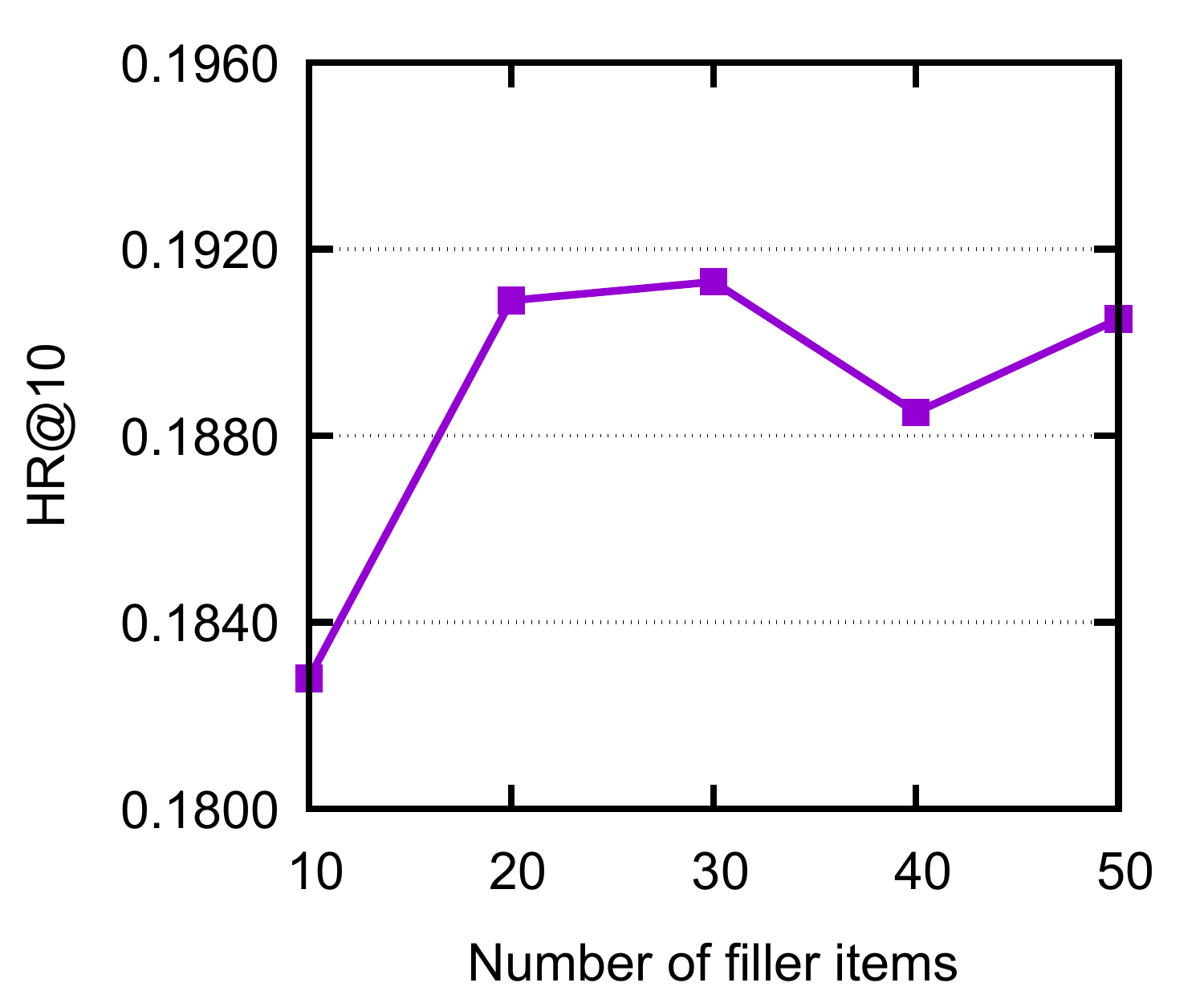}
		\caption{Video}
	\end{subfigure}
	\caption{Impact of the number of filler items.} \label{filler_para}
\end{figure}

\myparatight{Impact of the number of filler items} Figure~\ref{filler_para} shows the impact of the number of filler items on our attacks for random target items. On the Movie dataset, the hit ratio decreases as the attacker uses more filler items. However, on the Video dataset, the hit ratio increases and fluctuates as more filler items are used. Therefore, the relationship between the hit ratio and the number of filler items heavily depends on datasets. We note that Mobasher et al.~\cite{mobasher2007toward} had similar observations for the average and bandwagon attacks. Intuitively, an attacker should be more powerful and achieve better hit ratios when using more filler items. Our results and previous study~\cite{mobasher2007toward} show that this intuition does not hold. Understanding such phenomena theoretically is an interesting future work.

\subsection{Transferring to Other Systems}

In the previous section, we assume that the attacker has a \emph{white-box} access to the target recommender system. In this section, we consider an attacker has a \emph{gray-box} and \emph{black-box} access to the recommender system. In particular, in the gray-box setting, the recommender system is still graph-based recommender system, but the key parameter \emph{restart probability $\alpha$} is unknown to the attacker. In the black-box setting, the attacker does not know the target recommender system algorithm. To simulate such black-box setting,  we assume the attacker generates fake users based on a graph-based recommender system, while the target recommender system uses matrix factorization. 

 \begin{figure}[!t]
	\centering
		\includegraphics[width=0.45\textwidth]{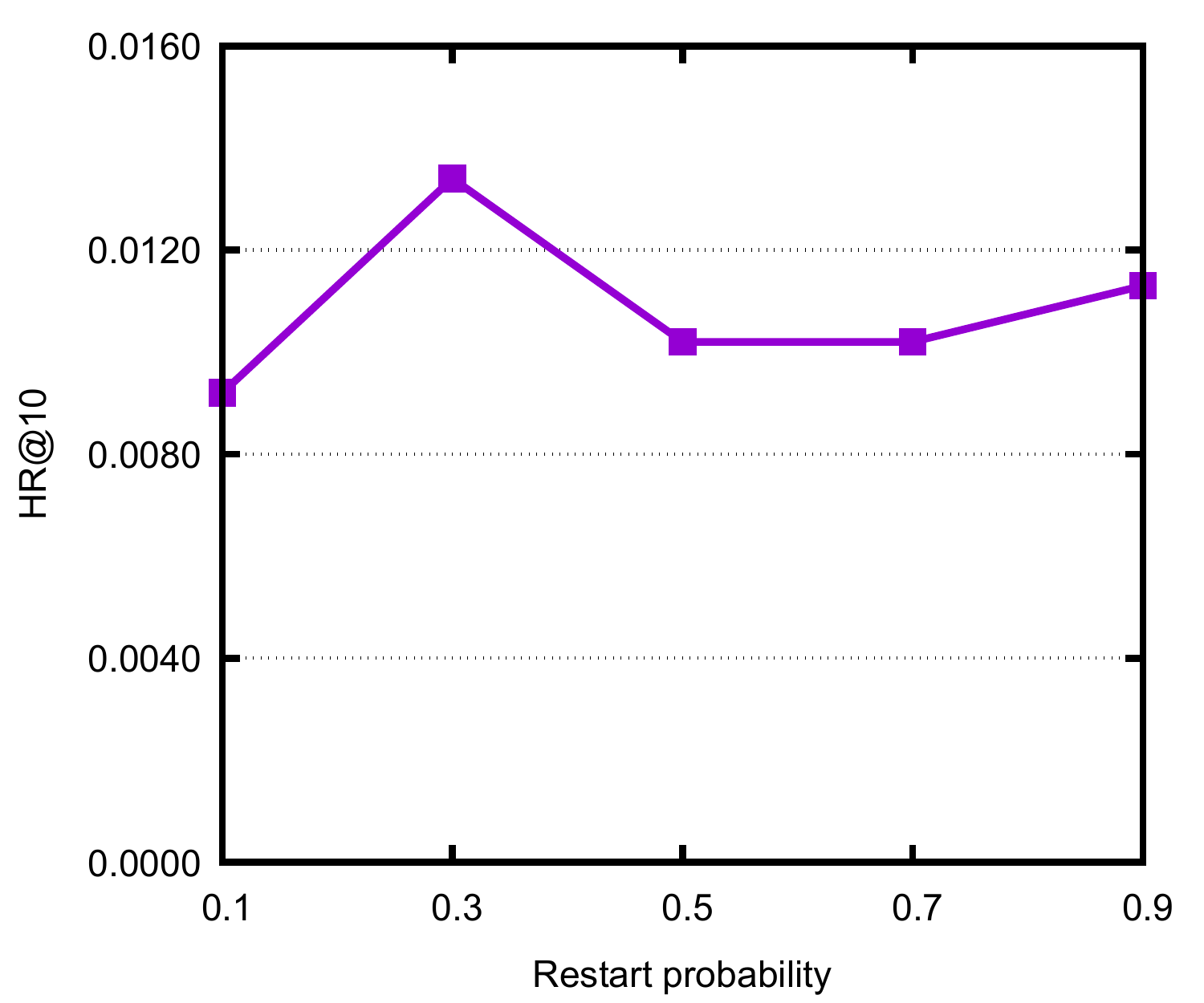}
	\caption{Hit ratio of our attack as a function of the restart probability of the target graph-based recommender system under the gray-box setting.} 
	\label{restart_para}
\end{figure}

\myparatight{Gray-box setting} The attacker uses a restart probability $\alpha=0.3$ in graph-based recommender system to generate fake users. Figure~\ref{restart_para} shows the hit ratios for random target items of our attacks when the target graph-based recommender system uses different restart probabilities.  We observe that the hit ratio reaches the maximum when the restart probability is 0.3. The reason is that the attacker also sets the restart probability to be 0.3, which essentially reduces to a white-box attack. When the target recommender system uses a 
restart probability other than 0.3, our attack is less effective. However, our attack is still much more effective than existing attacks (please refer to Table~\ref{Data_poisoning_attacks_Results}). 


\myparatight{Black-box setting} We assume the attacker generates fake users using a graph-based recommender system, while the target recommender system uses matrix factorization.  In particular, we use the popular matrix factorization technique proposed in~\cite{MFRec09} to implement the target recommender system. Table~\ref{MF_attack} shows the hit ratios of our attacks and existing attacks for random target items. First, all compared attacks can transfer to matrix factorization based recommender systems, especially on the Video dataset. Specifically, all attacks significantly improve the hit ratios of target items upon no attacks on the Video dataset. However, the hit ratio gains on the Movie dataset is less significant. We suspect the reason is that the Movie dataset is denser and is harder to attack. 

Second, the differences between our attack and the existing attacks are small, which means that different attacks have similar \emph{transferability} to matrix factorization based recommender systems.   
Third, the hit ratio gains of all attacks are less (or more) significant in the black-box setting than in the white-box setting on the Movie (or Video) dataset (comparing Table~\ref{Data_poisoning_attacks_Results} and Table~\ref{MF_attack}). For instance, on the Movie dataset, our attack improves the hit ratio over no attacks by 3 times and by 20\%  in the white-box setting and black-box setting, respectively, when the attack size is 1\%. However, on the Video dataset, our attack improves the hit ratio over no attacks by 33 times and 4000 times  in the white-box setting and black-box setting, respectively, when the attack size is 1\%. This is because matrix factorization is known to achieve better hit ratios when the dataset is denser~\cite{MFRec09}. For instance, matrix factorization achieves lower hit ratios than the graph-based recommender system on the Video dataset when there are no attacks. After the attacker adds fake users, the target item has dense rating scores and thus it is recommended to many users by matrix factorization. As a result, the poisoning attacks have even more significant hit ratio gains over no attacks in the black-box setting than in the white-box setting.

\begin{table}[!t]
	\centering
	\caption{HR@10 under the black-box setting, where the attacker generates fake users using a graph-based recommender system while the target recommender system uses matrix factorization.}\label{MF_attack}
	\begin{tabular}{c|c|cccc}
		\hline
		\multirow{2}[4]{*}{Dataset} & \multirow{2}[4]{*}{Attack } & \multicolumn{4}{c}{Attack size} \bigstrut\\
		\cline{3-6}          &       & 0.5\% & 1\%   & 3\%   & 5\% \bigstrut\\
		\hline
		\hline
		\multirow{6}[1]{*}{Movie} & None  & 0.0104 & 0.0104 & 0.0104 & 0.0104 \bigstrut[t]\\
		& Random & \textbf{0.0116} & \textbf{0.0125} & 0.0144 & 0.0198 \\
		& Average & \textbf{0.0116} & \textbf{0.0125} & 0.0144 & 0.0196 \\
		& Bandwagon & \textbf{0.0116} & \textbf{0.0125} & 0.0144 & 0.0198 \\
		& Co-visitation & 0.0015 & \textbf{0.0125} & 0.0144 & 0.0196 \\
		& Our attack & \textbf{0.0116} & 0.0124 & \textbf{0.0169} & \textbf{0.0226} \\
		\hline
		\multirow{6}[1]{*}{Video} & None  & 0.0001 & 0.0001 & 0.0001 & 0.0001 \\
		& Random & 0.0336 & 0.4142 & 0.5771 & 0.5884 \\
		& Average & 0.0317 & 0.4149 & \textbf{0.5776} & \textbf{0.5895} \\
		& Bandwagon & 0.0329 & 0.4142 & 0.5773 & 0.5883 \\
		& Co-visitation & 0.0325 & 0.4145 & 0.5775 & 0.5886 \\
		& Our attack & \textbf{0.0340} & \textbf{0.4158} & 0.5767 & 0.5852 \bigstrut[b]\\
		\hline
		\hline
	\end{tabular}%
\end{table}%


\begin{table*}[ht]
	\centering
	\caption{Detection results for different attacks.}\label{Detection_resultsl}
	\begin{tabular}{c|c|cccc|cccc}
		\hline
		\multirow{2}[4]{*}{Dataset} & \multirow{2}[4]{*}{Attack} & \multicolumn{4}{c|}{FPR}       & \multicolumn{4}{c}{FNR} \bigstrut\\
		\cline{3-10}          &       & 0.5\% & 1\%   & 3\%   & 5\%   & 0.5\% & 1\%   & 3\%   & 5\% \bigstrut\\
		\hline
		\hline
		\multirow{5}[2]{*}{Movie} & Random & 0.0458 & \textbf{0.0477} & 0.0463 & 0.0483 & 0     & 0.2400 & 0.4367 & 0.4060 \bigstrut[t]\\
		& Average & 0.0468 & 0.0463 & 0.0478 & 0.0475 & 0     & 0.2300 & \textbf{0.4567} & 0.4140 \\
		& Bandwagon & {0.0414} & {0.0417} & {0.0415} & {0.0445} & 0     & 0.1900 & 0.4100 & 0.3400 \\
		& Co-visitation & 0.0437 & 0.0460 & 0.0454 & 0.0461 & 0     & 0.2400 & 0.4200 & 0.3950 \\
		& Our attack & \textbf{0.0479} & 0.0474 & \textbf{0.0483} & \textbf{0.0493} & \textbf{0.1600} & \textbf{0.3900} & \textbf{0.4567} & \textbf{0.4220} \bigstrut[b]\\
		\hline
		\multirow{5}[2]{*}{Video} & Random & {0.0719} & 0.0717 & 0.0706 & 0.0719 & \textbf{0.2520} & 0.2820 & 0.3773 & 0.5116 \bigstrut[t]\\
		& Average & 0.0721 & 0.0700 & {0.0698} & {0.0693} & 0.2120 & \textbf{0.2860} & \textbf{0.3820} & \textbf{0.5452} \\
		& Bandwagon & 0.0721 & {0.0699} & 0.0701 & 0.0702 & 0.2040 & 0.2740 & 0.3467 & 0.5404 \\
		& Co-visitation & {0.0719} & 0.0705 & 0.0705 & 0.0702 & 0.2105 & 0.2850 & 0.3747 & 0.5369 \\
		& Our attack & \textbf{0.0730} & \textbf{0.0729} & \textbf{0.0725} & \textbf{0.0725} & 0.1880 & 0.2220 & 0.3353 & 0.5000 \bigstrut[b]\\
		\hline
		\hline
	\end{tabular}%
\end{table*}%

\begin{table}[ht]
	\centering
	\caption{HR@10 for different attacks when the service provider deploys a classifier to predict fake users and excludes the predicted fake users from the  system.}\label{attackUnderDetection}
	\begin{tabular}{c|c|cccc}
		\hline
		\multirow{2}[4]{*}{Dataset} & \multirow{2}[4]{*}{Attack } & \multicolumn{4}{c}{Attack size} \bigstrut\\
		\cline{3-6}          &       & 0.5\% & 1\%   & 3\%   & 5\% \bigstrut\\
		\hline
		\hline
		\multirow{6}[1]{*}{Movie} & None  & 0     & 0     & 0     & 0 \bigstrut[t]\\
		& Random & 0     & 0     & 0.0011 & 0.0031 \\
		& Average & 0     & 0     & 0.0014 & 0.0032 \\
		& Bandwagon & 0     & 0     & 0     & 0.0016 \\
		& Co-visitation & 0     & 0     & 0.0009 & 0.0027 \\
		& Our attack & 0     & \textbf{0.0024} & \textbf{0.0066} & \textbf{0.0109} \\
		\hline
		\multirow{6}[1]{*}{Video} & None  & 0.0008 & 0.0008 & 0.0008 & 0.0008 \\
		& Random & 0.0046 & 0.0087 & 0.0713 & 0.1530 \\
		& Average & 0.0058 & 0.0098 & 0.0891 & 0.1509 \\
		& Bandwagon & 0.0032 & 0.0071 & 0.0772 & 0.1459 \\
		& Co-visitation & 0.0025 & 0.0077 & 0.0739 & 0.1515 \\
		& Our attack & \textbf{0.0064} & \textbf{0.0197} & \textbf{0.1028} & \textbf{0.1788} \bigstrut[b]\\
		\hline
		\hline
	\end{tabular}%
\end{table}%

\section{Detecting Fake Users}
Detecting fake users is closely related to \emph{Sybil detection} in social networks. Many methods have been developed for Sybil detection. 
These methods leverage IP addresses (e.g.,~\cite{EVILCOHORTSecurity15}), user behavior (e.g.,~\cite{Wang13Clickstream}), or social relationships between users (e.g.,~\cite{sybilbelief,wang2017gang,wang2018structure}). Since we do not have access to IP addresses nor social relationships of users, we explore a behavior based method. 
In particular, we extract a set of features from a user's rating scores and train a binary classifier to classify users to be normal or fake. We will also study the effectiveness of the poisoning attacks when the recommender system has deployed such a detector to predict fake users and has excluded the predicted fake users from the recommender system. 

\myparatight{Rating scores based detection} Intuitively, the fake users' rating scores are generated in specific ways, and thus it may be possible to distinguish between normal users and fake users using their rating scores. Indeed, previous studies~\cite{chirita2005preventing,mobasher2007toward} extracted several features from rating scores to train a binary classifier to distinguish between normal users and fake users. We adopt these features in our work. Specifically, the features are as follows.


\begin{itemize}
	\item \textit{Rating Deviation from Mean Agreement} (RDMA)~\cite{chirita2005preventing}: This feature measures the average deviation of a user's rating scores to the mean rating scores of the corresponding items. Formally, for a user $u$, RDMA is computed as follows:
	\begin{align}
	RDM{A_u} = \frac{{\sum\limits_{i \in {I_u}} {\frac{{|{r_{ui}} - \overline {r_i}|}}{{{o_i}}}}}}{{|{I_u}|}},
	\end{align}
	where ${{I_u}}$ is the set of items that user $u$ has rated, ${\left| {{I_u}} \right|}$ is the number of items in ${{I_u}}$, $r_{ui}$ is the rating score that $u$ gave to item $i$, $\overline {r_i}$ is the average rating score for item $i$, and $o_i$ is the total number of ratings  for item $i$.
	
	\item \textit{Weighted Degree of Agreement} (WDA)~\cite{mobasher2007toward}: This feature is simply the numerator of the RDMA feature, i.e., this feature is computed as follows:
	\begin{align}
	WD{A_u} = \sum\limits_{i \in {I_u}} {\frac{{|{r_{ui}} - \overline {r_i}|}}{{{o_i}}}}.
	\end{align}
	\item \textit{Weighted Deviation from Mean Agreement} (WDMA)~\cite{mobasher2007toward}: This feature is also based on RDMA, but it puts higher weights on rating deviations for items that have less ratings. The WDMA feature for a user $u$ is calculated as follows:
	\begin{align}
	 WDM{A_u} = \frac{{\sum\limits_{i \in {I_u}} {\frac{{|{r_{ui}} - \overline {r_i}|}}{{o_i^2}}} }}{{|{I_u}|}}.
	\end{align}
	
	\item \textit{Mean Variance} (MeanVar)~\cite{mobasher2007toward}: This feature measures the average variance of a user's rating scores to the mean rating scores of the corresponding items. Specifically,  the MeanVar feature for a user $u$ is given by:
	\begin{align}
	MeanVa{r_u} = \frac{{\sum\limits_{i \in {I_{u}}} {{{({r_{ui}} - \overline {r_i})}^2}} }}{{|{I_u}|}}.
	\end{align}
	
	\item \textit{Filler Mean Target Difference} (FMTD)~\cite{mobasher2007toward}: This feature measures the divergence between a user's rating scores, and it is computed as follows:
	\begin{align}
	FMT{D_u} = \left| {\frac{{\sum\limits_{i \in {I_{uT}}} {{r_{ui}}} }}{{|{I_{uT}}|}} - \frac{{\sum\limits_{j \in {I_{uF}}} {{r_{uj}}} }}{{|{I_{uF}}|}}} \right|,
	\end{align}
	where ${I_{uT}}$ is the set of items in $I_u$ that $u$ gave the maximum rating score and ${I_{uF}}$ includes all other items in $I_u$.
	
\end{itemize}

For each poisoning attack, the service provider generates some fake users using the attack and labels some normal users as a training dataset. In our experiments, we generate 150 fake users (these fake users could be different from the fake users an attacker synthesizes when performing attacks) and sample 150 normal users as the training dataset. Then, using the above features, the service provider learns a KNN classifier, where $K$ is determined via cross-validation in the training dataset. 


\myparatight{Results of detecting fake users} We apply the classifiers to detect fake users generated by different poisoning attacks. We use \emph{False Positive Rate (FPR)} and \emph{False Negative Rate (FNR)} to measure the detection performance. Specifically, FPR is the fraction of normal users that are predicted to be fake, while FNR is the fraction of fake users that are predicted to be normal.

Table~\ref{Detection_resultsl} shows the FPR and FNR when detecting the fake users generated in our experiments in Section~\ref{sec:exp} under the default parameter setting. We observe that a small fraction of normal users are predicted to be fake. When the service provider excludes the predicted fake users from the recommender system, these normal users won't receive \emph{personalized} recommendations from the recommender system. The service provider could leverage other methods to recommend items for such users, e.g., the service provider always recommends popular items to them (such recommendation is not personalized). Moreover, the detector misses a large fraction of fake users, i.e., FNR is large. Moreover, the FNR tends to increase as the attacker injects more fake users. A possible reason is that more fake users have more diverse patterns, and thus it is harder to detect them.


\myparatight{Attack effectiveness when detector is deployed} Suppose the service provider deploys the classifier to detect fake users. In particular, the service provider excludes the predicted fake users from the recommender system. Note that a small fraction of normal users will be excluded from the recommender system, while a large fraction of fake users will still be included in the recommender system. We re-compute the recommended items for each remaining user after excluding the predicted fake users from the recommender system and re-compute the hit ratios of the target items. The hit ratio of a target item is the fraction of the remaining normal users whose recommended items include the target item. 

Table~\ref{attackUnderDetection} shows the hit ratios of random target items for the compared poisoning attacks under the white-box setting.  First, we observe that these attacks are still effective in many cases. This is because a large fraction of fake users are not detected. Second, compared to the case where the service provider does not detect fake users, the hit ratios  are smaller (comparing Table~\ref{attackUnderDetection} with Table~\ref{Data_poisoning_attacks_Results}). The reason is that a large fraction of fake users are detected and excluded from the recommender system.  Third, our attack still substantially outperforms existing attacks. 

\section{Conclusion and Future Work} \label{sec:conclusion}

In this work, we propose optimized poisoning attacks to graph-based recommender systems. We show that poisoning attacks to graph-based recommender systems can be formulated as an optimization problem and the optimization problem can be approximately solved by a projected gradient descent method. Via evaluations on real-world datasets, we find that our attacks can make a target item recommended to substantially more users. Moreover, our attacks are more effective than existing attacks for manipulating graph-based recommender systems. The reason is that existing attacks are not optimized for graph-based recommender systems, while our attacks are. Our attacks can also transfer to other recommender systems under the gray-box and black-box settings. The service provider can detect a large fraction of fake users but also falsely predict a small fraction of normal users to be fake, via using supervised machine learning techniques to analyze the users' rating scores. Moreover, our attacks are still effective when the service provider deploys such a detector and excludes the predicted fake users from the recommender system. 

Interesting future works include 1) evaluating our poisoning attacks on real-world graph-based recommender systems, 2) designing optimized poisoning attacks to other graph-based recommender systems (e.g., graph convolutional neural network based recommender systems~\cite{Ying18}), 3) designing optimized poisoning attacks to neural network based recommender systems (e.g.,~\cite{He17}), and 4) designing defenses against poisoning attacks.

\section*{Acknowledgements}
We thank the anonymous reviewers for their insightful feedback. This work is supported by the National Science Foundation under Grants No.
 CNS-1750198 and CNS-1801584. Any opinions, findings and conclusions or recommendations
expressed in this material are those of the author(s) and do
not necessarily reflect the views of the funding agencies.


\balance
\bibliographystyle{ACM-Reference-Format}
\bibliography{mdb,ref,refs}


\begin{thebibliography}{37}


\ifx \showCODEN    \undefined \def \showCODEN     #1{\unskip}     \fi
\ifx \showDOI      \undefined \def \showDOI       #1{#1}\fi
\ifx \showISBNx    \undefined \def \showISBNx     #1{\unskip}     \fi
\ifx \showISBNxiii \undefined \def \showISBNxiii  #1{\unskip}     \fi
\ifx \showISSN     \undefined \def \showISSN      #1{\unskip}     \fi
\ifx \showLCCN     \undefined \def \showLCCN      #1{\unskip}     \fi
\ifx \shownote     \undefined \def \shownote      #1{#1}          \fi
\ifx \showarticletitle \undefined \def \showarticletitle #1{#1}   \fi
\ifx \showURL      \undefined \def \showURL       {\relax}        \fi
\providecommand\bibfield[2]{#2}
\providecommand\bibinfo[2]{#2}
\providecommand\natexlab[1]{#1}
\providecommand\showeprint[2][]{arXiv:#2}

\bibitem[\protect\citeauthoryear{??}{Cam}{2018}]%
        {Cambridge}
 \bibinfo{year}{2018}\natexlab{}.
\newblock \bibinfo{title}{Cambridge Analytica}.
\newblock
\newblock
\urldef\tempurl%
\url{https://www.theguardian.com/technology/2018/mar/17/facebook-cambridge-analytica-kogan-data-algorithm}
\showURL{%
\tempurl}


\bibitem[\protect\citeauthoryear{{Amazon Instant Video Dataset.}}{{Amazon
  Instant Video Dataset.}}{2018}]%
        {amazonURL}
\bibfield{author}{\bibinfo{person}{{Amazon Instant Video Dataset.}}}
  \bibinfo{year}{2018}\natexlab{}.
\newblock
\newblock
\urldef\tempurl%
\url{http://jmcauley.ucsd.edu/data/amazon/}
\showURL{%
\tempurl}


\bibitem[\protect\citeauthoryear{Backstrom and Leskovec}{Backstrom and
  Leskovec}{2011}]%
        {backstrom2011supervised}
\bibfield{author}{\bibinfo{person}{Lars Backstrom} {and} \bibinfo{person}{Jure
  Leskovec}.} \bibinfo{year}{2011}\natexlab{}.
\newblock \showarticletitle{Supervised random walks: predicting and
  recommending links in social networks}. In
  \bibinfo{booktitle}{\emph{Proceedings of the fourth ACM international
  conference on Web search and data mining}}. ACM, \bibinfo{pages}{635--644}.
\newblock


\bibitem[\protect\citeauthoryear{Calandrino, Kilzer, Narayanan, Felten, and
  Shmatikov}{Calandrino et~al\mbox{.}}{2011}]%
        {CalandrinoOakland11}
\bibfield{author}{\bibinfo{person}{Joseph~A. Calandrino}, \bibinfo{person}{Ann
  Kilzer}, \bibinfo{person}{Arvind Narayanan}, \bibinfo{person}{Edward~W.
  Felten}, {and} \bibinfo{person}{Vitaly Shmatikov}.}
  \bibinfo{year}{2011}\natexlab{}.
\newblock \showarticletitle{``You Might Also Like:'' Privacy Risks of
  Collaborative Filtering}. In \bibinfo{booktitle}{\emph{IEEE Symposium on
  Security and Privacy}}.
\newblock


\bibitem[\protect\citeauthoryear{Chirita, Nejdl, and Zamfir}{Chirita
  et~al\mbox{.}}{2005}]%
        {chirita2005preventing}
\bibfield{author}{\bibinfo{person}{Paul-Alexandru Chirita},
  \bibinfo{person}{Wolfgang Nejdl}, {and} \bibinfo{person}{Cristian Zamfir}.}
  \bibinfo{year}{2005}\natexlab{}.
\newblock \showarticletitle{Preventing shilling attacks in online recommender
  systems}. In \bibinfo{booktitle}{\emph{Proceedings of the 7th annual ACM
  international workshop on Web information and data management}}. ACM,
  \bibinfo{pages}{67--74}.
\newblock


\bibitem[\protect\citeauthoryear{Davidson, Liebald, Liu, Nandy, Van~Vleet,
  Gargi, Gupta, He, Lambert, Livingston, et~al\mbox{.}}{Davidson
  et~al\mbox{.}}{2010}]%
        {davidson2010youtube}
\bibfield{author}{\bibinfo{person}{James Davidson}, \bibinfo{person}{Benjamin
  Liebald}, \bibinfo{person}{Junning Liu}, \bibinfo{person}{Palash Nandy},
  \bibinfo{person}{Taylor Van~Vleet}, \bibinfo{person}{Ullas Gargi},
  \bibinfo{person}{Sujoy Gupta}, \bibinfo{person}{Yu He}, \bibinfo{person}{Mike
  Lambert}, \bibinfo{person}{Blake Livingston}, {et~al\mbox{.}}}
  \bibinfo{year}{2010}\natexlab{}.
\newblock \showarticletitle{The YouTube video recommendation system}. In
  \bibinfo{booktitle}{\emph{ACM conference on Recommender systems}}. ACM,
  \bibinfo{pages}{293--296}.
\newblock


\bibitem[\protect\citeauthoryear{Fouss, Pirotte, Renders, and Saerens}{Fouss
  et~al\mbox{.}}{2007}]%
        {fouss2007random}
\bibfield{author}{\bibinfo{person}{Francois Fouss}, \bibinfo{person}{Alain
  Pirotte}, \bibinfo{person}{Jean-Michel Renders}, {and} \bibinfo{person}{Marco
  Saerens}.} \bibinfo{year}{2007}\natexlab{}.
\newblock \showarticletitle{Random-walk computation of similarities between
  nodes of a graph with application to collaborative recommendation}.
\newblock \bibinfo{journal}{\emph{IEEE Transactions on knowledge and data
  engineering}} \bibinfo{volume}{19}, \bibinfo{number}{3}
  (\bibinfo{year}{2007}), \bibinfo{pages}{355--369}.
\newblock


\bibitem[\protect\citeauthoryear{Gong, Frank, and Mittal}{Gong
  et~al\mbox{.}}{2014a}]%
        {sybilbelief}
\bibfield{author}{\bibinfo{person}{Neil~Zhenqiang Gong}, \bibinfo{person}{Mario
  Frank}, {and} \bibinfo{person}{Prateek Mittal}.}
  \bibinfo{year}{2014}\natexlab{a}.
\newblock \showarticletitle{SybilBelief: {A} Semi-supervised Learning Approach
  for Structure-based Sybil Detection}.
\newblock \bibinfo{journal}{\emph{IEEE Transactions on Information Forensics
  and Security}} \bibinfo{volume}{9}, \bibinfo{number}{6}
  (\bibinfo{year}{2014}), \bibinfo{pages}{976 -- 987}.
\newblock


\bibitem[\protect\citeauthoryear{Gong and Liu}{Gong and Liu}{2016}]%
        {GongAttriInferSEC16}
\bibfield{author}{\bibinfo{person}{Neil~Zhenqiang Gong} {and}
  \bibinfo{person}{Bin Liu}.} \bibinfo{year}{2016}\natexlab{}.
\newblock \showarticletitle{You are Who You Know and How You Behave: Attribute
  Inference Attacks via Users' Social Friends and Behaviors}. In
  \bibinfo{booktitle}{\emph{USENIX Security Symposium}}.
\newblock


\bibitem[\protect\citeauthoryear{Gong and Liu}{Gong and Liu}{2018}]%
        {gongTOPS18}
\bibfield{author}{\bibinfo{person}{Neil~Zhenqiang Gong} {and}
  \bibinfo{person}{Bin Liu}.} \bibinfo{year}{2018}\natexlab{}.
\newblock \showarticletitle{Attribute Inference Attacks in Online Social
  Networks}.
\newblock \bibinfo{journal}{\emph{ACM Transactions on Privacy and Security
  (TOPS)}} \bibinfo{volume}{21}, \bibinfo{number}{1} (\bibinfo{year}{2018}).
\newblock


\bibitem[\protect\citeauthoryear{Gong, Talwalkar, Mackey, Huang, Shin,
  Stefanov, Shi, and Song}{Gong et~al\mbox{.}}{2014b}]%
        {Gong14}
\bibfield{author}{\bibinfo{person}{Neil~Zhenqiang Gong}, \bibinfo{person}{Ameet
  Talwalkar}, \bibinfo{person}{Lester Mackey}, \bibinfo{person}{Ling Huang},
  \bibinfo{person}{Eui Chul~Richard Shin}, \bibinfo{person}{Emil Stefanov},
  \bibinfo{person}{Elaine(Runting) Shi}, {and} \bibinfo{person}{Dawn Song}.}
  \bibinfo{year}{2014}\natexlab{b}.
\newblock \showarticletitle{Joint Link Prediction and Attribute Inference using
  a Social-Attribute Network}.
\newblock \bibinfo{journal}{\emph{ACM TIST}} \bibinfo{volume}{5},
  \bibinfo{number}{2} (\bibinfo{year}{2014}).
\newblock


\bibitem[\protect\citeauthoryear{Guo, Tang, Ye, Li, and He}{Guo
  et~al\mbox{.}}{2017}]%
        {Guo17}
\bibfield{author}{\bibinfo{person}{Huifeng Guo}, \bibinfo{person}{Ruiming
  Tang}, \bibinfo{person}{Yunming Ye}, \bibinfo{person}{Zhenguo Li}, {and}
  \bibinfo{person}{Xiuqiang He}.} \bibinfo{year}{2017}\natexlab{}.
\newblock \showarticletitle{A Graph-based Push Service Platform}. In
  \bibinfo{booktitle}{\emph{DASFAA}}.
\newblock


\bibitem[\protect\citeauthoryear{He, Dai, Cao, Tang, Yuan, and Yang}{He
  et~al\mbox{.}}{2015}]%
        {He15}
\bibfield{author}{\bibinfo{person}{Xiuqiang He}, \bibinfo{person}{Wenyuan Dai},
  \bibinfo{person}{Guoxiang Cao}, \bibinfo{person}{Ruiming Tang},
  \bibinfo{person}{Mingxuan Yuan}, {and} \bibinfo{person}{Qiang Yang}.}
  \bibinfo{year}{2015}\natexlab{}.
\newblock \showarticletitle{Mining target users for online marketing based on
  App Store data}. In \bibinfo{booktitle}{\emph{IEEE International Conference
  on Big Data}}.
\newblock


\bibitem[\protect\citeauthoryear{He, Liao, Zhang, Nie, Hu, and Chua}{He
  et~al\mbox{.}}{2017}]%
        {He17}
\bibfield{author}{\bibinfo{person}{Xiangnan He}, \bibinfo{person}{Lizi Liao},
  \bibinfo{person}{Hanwang Zhang}, \bibinfo{person}{Liqiang Nie},
  \bibinfo{person}{Xia Hu}, {and} \bibinfo{person}{Tat-Seng Chua}.}
  \bibinfo{year}{2017}\natexlab{}.
\newblock \showarticletitle{Neural Collaborative Filtering}. In
  \bibinfo{booktitle}{\emph{WWW}}.
\newblock


\bibitem[\protect\citeauthoryear{Jia and Gong}{Jia and Gong}{2018}]%
        {Jia18}
\bibfield{author}{\bibinfo{person}{Jinyuan Jia} {and}
  \bibinfo{person}{Neil~Zhenqiang Gong}.} \bibinfo{year}{2018}\natexlab{}.
\newblock \showarticletitle{AttriGuard: A Practical Defense Against Attribute
  Inference Attacks via Adversarial Machine Learning}. In
  \bibinfo{booktitle}{\emph{USENIX Security Symposium}}.
\newblock


\bibitem[\protect\citeauthoryear{Jia, Wang, Zhang, and Gong}{Jia
  et~al\mbox{.}}{2017}]%
        {AttriInfer}
\bibfield{author}{\bibinfo{person}{Jinyuan Jia}, \bibinfo{person}{Binghui
  Wang}, \bibinfo{person}{Le Zhang}, {and} \bibinfo{person}{Neil~Zhenqiang
  Gong}.} \bibinfo{year}{2017}\natexlab{}.
\newblock \showarticletitle{{AttriInfer}: Inferring User Attributes in Online
  Social Networks Using Markov Random Fields}. In
  \bibinfo{booktitle}{\emph{WWW}}.
\newblock


\bibitem[\protect\citeauthoryear{Koren, Bell, and Volinsky}{Koren
  et~al\mbox{.}}{2009}]%
        {MFRec09}
\bibfield{author}{\bibinfo{person}{Y Koren}, \bibinfo{person}{R Bell}, {and}
  \bibinfo{person}{C Volinsky}.} \bibinfo{year}{2009}\natexlab{}.
\newblock \showarticletitle{Matrix factorization techniques for recommender
  systems}.
\newblock \bibinfo{journal}{\emph{Computer}}  \bibinfo{volume}{8}
  (\bibinfo{year}{2009}), \bibinfo{pages}{30--37}.
\newblock


\bibitem[\protect\citeauthoryear{Kosinski, Stillwell, and Graepel}{Kosinski
  et~al\mbox{.}}{2013}]%
        {kosinski2013private}
\bibfield{author}{\bibinfo{person}{Michal Kosinski}, \bibinfo{person}{David
  Stillwell}, {and} \bibinfo{person}{Thore Graepel}.}
  \bibinfo{year}{2013}\natexlab{}.
\newblock \showarticletitle{Private traits and attributes are predictable from
  digital records of human behavior}.
\newblock \bibinfo{journal}{\emph{PNAS}} (\bibinfo{year}{2013}).
\newblock


\bibitem[\protect\citeauthoryear{Lam and Riedl}{Lam and Riedl}{2004}]%
        {lam2004shilling}
\bibfield{author}{\bibinfo{person}{Shyong~K Lam} {and} \bibinfo{person}{John
  Riedl}.} \bibinfo{year}{2004}\natexlab{}.
\newblock \showarticletitle{Shilling recommender systems for fun and profit}.
  In \bibinfo{booktitle}{\emph{WWW}}.
\newblock


\bibitem[\protect\citeauthoryear{Li, Wang, Singh, and Vorobeychik}{Li
  et~al\mbox{.}}{2016}]%
        {poisoningattackRecSys16}
\bibfield{author}{\bibinfo{person}{Bo Li}, \bibinfo{person}{Yining Wang},
  \bibinfo{person}{Aarti Singh}, {and} \bibinfo{person}{Yevgeniy Vorobeychik}.}
  \bibinfo{year}{2016}\natexlab{}.
\newblock \showarticletitle{Data Poisoning Attacks on Factorization-Based
  Collaborative Filtering}. In \bibinfo{booktitle}{\emph{NIPS}}.
\newblock


\bibitem[\protect\citeauthoryear{Mobasher, Burke, Bhaumik, and
  Williams}{Mobasher et~al\mbox{.}}{2007}]%
        {mobasher2007toward}
\bibfield{author}{\bibinfo{person}{Bamshad Mobasher}, \bibinfo{person}{Robin
  Burke}, \bibinfo{person}{Runa Bhaumik}, {and} \bibinfo{person}{Chad
  Williams}.} \bibinfo{year}{2007}\natexlab{}.
\newblock \showarticletitle{Toward trustworthy recommender systems: An analysis
  of attack models and algorithm robustness}.
\newblock \bibinfo{journal}{\emph{ACM Transactions on Internet Technology}}
  \bibinfo{volume}{7}, \bibinfo{number}{4} (\bibinfo{year}{2007}),
  \bibinfo{pages}{23}.
\newblock


\bibitem[\protect\citeauthoryear{Mobasher, Cooley, and Srivastava}{Mobasher
  et~al\mbox{.}}{2000}]%
        {mobasher2000automatic}
\bibfield{author}{\bibinfo{person}{Bamshad Mobasher}, \bibinfo{person}{Robert
  Cooley}, {and} \bibinfo{person}{Jaideep Srivastava}.}
  \bibinfo{year}{2000}\natexlab{}.
\newblock \showarticletitle{Automatic personalization based on web usage
  mining}.
\newblock \bibinfo{journal}{\emph{Commun. ACM}} \bibinfo{volume}{43},
  \bibinfo{number}{8} (\bibinfo{year}{2000}), \bibinfo{pages}{142--151}.
\newblock


\bibitem[\protect\citeauthoryear{{MovieLens Dataset.}}{{MovieLens
  Dataset.}}{2018}]%
        {movielensURL}
\bibfield{author}{\bibinfo{person}{{MovieLens Dataset.}}}
  \bibinfo{year}{2018}\natexlab{}.
\newblock
\newblock
\urldef\tempurl%
\url{https://grouplens.org/datasets/movielens/}
\showURL{%
\tempurl}


\bibitem[\protect\citeauthoryear{O'Mahony, Hurley, Kushmerick, and
  Silvestre}{O'Mahony et~al\mbox{.}}{2004}]%
        {profileinjectionattacksMahony04}
\bibfield{author}{\bibinfo{person}{M. O'Mahony}, \bibinfo{person}{N. Hurley},
  \bibinfo{person}{N. Kushmerick}, {and} \bibinfo{person}{G. Silvestre}.}
  \bibinfo{year}{2004}\natexlab{}.
\newblock \showarticletitle{Collaborative Recommendation: A Robustness
  Analysis}.
\newblock \bibinfo{journal}{\emph{ACM Transactions on Internet Technology}}
  \bibinfo{volume}{4}, \bibinfo{number}{4} (\bibinfo{year}{2004}),
  \bibinfo{pages}{344--377}.
\newblock


\bibitem[\protect\citeauthoryear{Pinckney}{Pinckney}{013a}]%
        {eBay2013a}
\bibfield{author}{\bibinfo{person}{Thomas Pinckney}.}
  \bibinfo{year}{2013a}\natexlab{}.
\newblock \bibinfo{title}{Graph-based Recommendation Systems at eBay}.
\newblock
\newblock
\urldef\tempurl%
\url{https://www.youtube.com/watch?t=2400&v=Tg3dP2fZGSM}
\showURL{%
\tempurl}


\bibitem[\protect\citeauthoryear{Pinckney}{Pinckney}{013b}]%
        {eBay2013b}
\bibfield{author}{\bibinfo{person}{Thomas Pinckney}.}
  \bibinfo{year}{2013b}\natexlab{}.
\newblock \bibinfo{title}{Planet Cassandra: Graph Based Recommendation Systems
  at eBay}.
\newblock
\newblock
\urldef\tempurl%
\url{http://www.slideshare.net/planetcassandra/e-bay-nyc}
\showURL{%
\tempurl}


\bibitem[\protect\citeauthoryear{Sarwar, Karypis, Konstan, and Riedl}{Sarwar
  et~al\mbox{.}}{2001}]%
        {sarwar2001item}
\bibfield{author}{\bibinfo{person}{Badrul Sarwar}, \bibinfo{person}{George
  Karypis}, \bibinfo{person}{Joseph Konstan}, {and} \bibinfo{person}{John
  Riedl}.} \bibinfo{year}{2001}\natexlab{}.
\newblock \showarticletitle{Item-based collaborative filtering recommendation
  algorithms}. In \bibinfo{booktitle}{\emph{Proceedings of the 10th
  international conference on World Wide Web}}. ACM, \bibinfo{pages}{285--295}.
\newblock


\bibitem[\protect\citeauthoryear{Stringhini, Mourlanne, Jacob, Egele, Kruegel,
  and Vigna}{Stringhini et~al\mbox{.}}{2015}]%
        {EVILCOHORTSecurity15}
\bibfield{author}{\bibinfo{person}{Gianluca Stringhini},
  \bibinfo{person}{Pierre Mourlanne}, \bibinfo{person}{Gregoire Jacob},
  \bibinfo{person}{Manuel Egele}, \bibinfo{person}{Christopher Kruegel}, {and}
  \bibinfo{person}{Giovanni Vigna}.} \bibinfo{year}{2015}\natexlab{}.
\newblock \showarticletitle{EVILCOHORT: Detecting Communities of Malicious
  Accounts on Online Services}. In \bibinfo{booktitle}{\emph{Usenix Security}}.
\newblock


\bibitem[\protect\citeauthoryear{Thomas, McCoy, Grier, Kolcz, and
  Paxson}{Thomas et~al\mbox{.}}{2013}]%
        {SybilUnderground13}
\bibfield{author}{\bibinfo{person}{Kurt Thomas}, \bibinfo{person}{Damon McCoy},
  \bibinfo{person}{Chris Grier}, \bibinfo{person}{Alek Kolcz}, {and}
  \bibinfo{person}{Vern Paxson}.} \bibinfo{year}{2013}\natexlab{}.
\newblock \showarticletitle{Trafficking Fraudulent Accounts: The Role of the
  Underground Market in Twitter Spam and Abuse}. In
  \bibinfo{booktitle}{\emph{USENIX Security Symposium}}.
\newblock


\bibitem[\protect\citeauthoryear{Wang, Gong, and Fu}{Wang
  et~al\mbox{.}}{2017}]%
        {wang2017gang}
\bibfield{author}{\bibinfo{person}{Binghui Wang},
  \bibinfo{person}{Neil~Zhenqiang Gong}, {and} \bibinfo{person}{Hao Fu}.}
  \bibinfo{year}{2017}\natexlab{}.
\newblock \showarticletitle{{GANG}: Detecting Fraudulent Users in Online Social
  Networks via Guilt-by-Association on Directed Graphs}. In
  \bibinfo{booktitle}{\emph{IEEE ICDM}}.
\newblock


\bibitem[\protect\citeauthoryear{Wang, Jia, Zhang, and Gong}{Wang
  et~al\mbox{.}}{2018}]%
        {wang2018structure}
\bibfield{author}{\bibinfo{person}{Binghui Wang}, \bibinfo{person}{Jinyuan
  Jia}, \bibinfo{person}{Le Zhang}, {and} \bibinfo{person}{Neil~Zhenqiang
  Gong}.} \bibinfo{year}{2018}\natexlab{}.
\newblock \showarticletitle{Structure-based Sybil Detection in Social Networks
  via Local Rule-based Propagation}.
\newblock \bibinfo{journal}{\emph{IEEE TNSE}} (\bibinfo{year}{2018}).
\newblock


\bibitem[\protect\citeauthoryear{Wang, Konolige, Wilson, and Wang}{Wang
  et~al\mbox{.}}{2013}]%
        {Wang13Clickstream}
\bibfield{author}{\bibinfo{person}{Gang Wang}, \bibinfo{person}{Tristan
  Konolige}, \bibinfo{person}{Christo Wilson}, {and} \bibinfo{person}{Xiao
  Wang}.} \bibinfo{year}{2013}\natexlab{}.
\newblock \showarticletitle{You are How You Click: Clickstream Analysis for
  Sybil Detection}. In \bibinfo{booktitle}{\emph{Usenix Security}}.
\newblock


\bibitem[\protect\citeauthoryear{Weinsberg, Bhagat, Ioannidis, and
  Taft}{Weinsberg et~al\mbox{.}}{2012}]%
        {weinsberg2012blurme}
\bibfield{author}{\bibinfo{person}{Udi Weinsberg}, \bibinfo{person}{Smriti
  Bhagat}, \bibinfo{person}{Stratis Ioannidis}, {and} \bibinfo{person}{Nina
  Taft}.} \bibinfo{year}{2012}\natexlab{}.
\newblock \showarticletitle{BlurMe: Inferring and obfuscating user gender based
  on ratings}. In \bibinfo{booktitle}{\emph{RecSys}}.
\newblock


\bibitem[\protect\citeauthoryear{Xing, Meng, Doozan, Snoeren, Feamster, and
  Lee}{Xing et~al\mbox{.}}{2013}]%
        {xing2013take}
\bibfield{author}{\bibinfo{person}{Xinyu Xing}, \bibinfo{person}{Wei Meng},
  \bibinfo{person}{Dan Doozan}, \bibinfo{person}{Alex~C Snoeren},
  \bibinfo{person}{Nick Feamster}, {and} \bibinfo{person}{Wenke Lee}.}
  \bibinfo{year}{2013}\natexlab{}.
\newblock \showarticletitle{Take This Personally: Pollution Attacks on
  Personalized Services}. In \bibinfo{booktitle}{\emph{USENIX Security}}.
  \bibinfo{pages}{671--686}.
\newblock


\bibitem[\protect\citeauthoryear{Yang, Gong, and Cai}{Yang
  et~al\mbox{.}}{2017}]%
        {yang2017fake}
\bibfield{author}{\bibinfo{person}{Guolei Yang},
  \bibinfo{person}{Neil~Zhenqiang Gong}, {and} \bibinfo{person}{Ying Cai}.}
  \bibinfo{year}{2017}\natexlab{}.
\newblock \showarticletitle{Fake Co-visitation Injection Attacks to Recommender
  Systems}. In \bibinfo{booktitle}{\emph{NDSS}}.
\newblock


\bibitem[\protect\citeauthoryear{Ying, He, Chen, Eksombatchai, Hamilton, and
  Leskovec}{Ying et~al\mbox{.}}{2018}]%
        {Ying18}
\bibfield{author}{\bibinfo{person}{Rex Ying}, \bibinfo{person}{Ruining He},
  \bibinfo{person}{Kaifeng Chen}, \bibinfo{person}{Pong Eksombatchai},
  \bibinfo{person}{William~L. Hamilton}, {and} \bibinfo{person}{Jure
  Leskovec}.} \bibinfo{year}{2018}\natexlab{}.
\newblock \showarticletitle{Graph Convolutional Neural Networks for Web-Scale
  Recommender Systems}. In \bibinfo{booktitle}{\emph{KDD}}.
\newblock


\bibitem[\protect\citeauthoryear{Zeller and Felten}{Zeller and Felten}{2008}]%
        {CSRF}
\bibfield{author}{\bibinfo{person}{William Zeller} {and}
  \bibinfo{person}{Edward~W. Felten}.} \bibinfo{year}{2008}\natexlab{}.
\newblock \bibinfo{booktitle}{\emph{Cross-Site Request Forgeries: Exploitation
  and Prevention}}.
\newblock \bibinfo{type}{{T}echnical {R}eport}. \bibinfo{institution}{Princeton
  University}.
\newblock


\end{thebibliography}

\end{document}